\documentclass{aa}
\usepackage{graphicx}
\usepackage{txfonts}
\usepackage{color}

\begin{document}
\title{{\it Hubble Space Telescope} imaging of the extremely metal-poor globular cluster EXT8 in Messier~31\thanks{Based on observations made with the NASA/ESA Hubble Space Telescope, obtained at the Space Telescope Science Institute, which is operated by the Association of Universities for Research in Astronomy, Inc., under NASA contract NAS5-26555. These observations are associated with program \#16459.}}

\author{
   S{\o}ren S. Larsen
   \inst{1}
   \and
     Aaron J. Romanowsky
   \inst{2,4}
   \and
    Jean P. Brodie
   \inst{3,4}
}

\institute{
  Department of Astrophysics/IMAPP,
              Radboud University, PO Box 9010, 6500 GL Nijmegen, The Netherlands\\
              \email{s.larsen@astro.ru.nl}
  \and
  Department of Physics \& Astronomy, One Washington Square, San Jos{\'e} State University, San Jose, CA 95192, USA
  \and
  Centre for Astrophysics and Supercomputing, Swinburne University of Technology, Hawthorn, VIC 3122, Australia
  \and
  University of California Observatories, 1156 High Street, Santa Cruz, CA 95064, USA
}

\date{Received 12 April 2021 / Accepted 4 May 2021}

\abstract
{We recently found the globular cluster (GC) EXT8 in M31 to have an extremely low metallicity of $\mathrm{[Fe/H]}=-2.91\pm0.04$ using high-resolution spectroscopy.
Here we present a colour--magnitude diagram (CMD) for EXT8, obtained with the Wide Field Camera~3 on board the {\it Hubble Space Telescope}.  Compared with the CMDs of metal-poor Galactic GCs, we find that the upper red giant branch (RGB) of EXT8 is $\sim0.03$ mag bluer in $M_\mathrm{F606W}-M_\mathrm{F814W}$ and slightly steeper, as expected from the low spectroscopic metallicity.
The observed colour spread on the upper RGB is consistent with being caused entirely by the measurement uncertainties, and we place an upper limit of $\sigma_\mathrm{F606W-F814W} \approx 0.015$~mag on any intrinsic colour spread. The corresponding metallicity spread can be up to 
$\sigma_\mathrm{[Fe/H]} \sim 0.2$~dex 
or $>0.7$~dex, depending on the isochrone library adopted.
The horizontal branch (HB) is located mostly on the blue side of the instability strip and has a tail extending to at least $M_\mathrm{F606W}=+3$, as in the Galactic GC M15. 
We identify two candidate RR Lyrae variables and several ultraviolet-luminous post-HB/post asymptotic giant branch star candidates, including one very bright ($M_\mathrm{F300X}\approx-3.2$) source near the centre of EXT8.
The surface brightness of EXT8 out to a radius of 25\arcsec\ is well fitted by a Wilson-type profile with an ellipticity of $\epsilon=0.20$, a semi-major axis core radius of 0\farcs25, and a central surface brightness of $\mu_{\mathrm{F606W},0} = 15.2$~mag~arcsec$^{-2}$, with no evidence of extra-tidal structure. Overall, EXT8 has properties consistent with it being a ``normal'', but very metal-poor GC, and its combination of relatively high mass and very low metallicity thus remains challenging to explain in the context of GC formation theories operating within the hierarchical galaxy assembly paradigm.
}


\titlerunning{Photometry of EXT8}
\authorrunning{S. Larsen et al.}
\maketitle

\section{Introduction}

In his investigation of globular cluster (GC) radial velocities, \citet{Mayall1946} noted that the integrated-light spectral types of GCs ranged from fairly early (A5) to roughly solar (G5). During the following decade, several authors linked the extremely weak metal lines in the spectra of GCs such as M15 and M92 to a deficiency in the abundances of the corresponding elements \citep{Baum1952,Morgan1956,Baade1958} and the first quantitative spectral analyses established that some GCs have metal/hydrogen ratios of less than 1\% of the solar value \citep{Helfer1959,Kinman1959}.
Measurements of chemical abundances and differences in the spatial distributions of metal-poor and metal-rich GCs were discussed in the context of the then-nascent fields of nucleosynthesis and Galactic chemical evolution \citep{Burbidge1957,Helfer1959,Morgan1959}, leading up to the seminal work of \citet{Searle1978}. Around the same time, classical investigations of GC colour--magnitude diagrams (CMDs) revealed correlations between metallicity and CMD characteristics such as the slope of the red giant branch (RGB) and the morphology of the horizontal branch (HB) \citep{Arp1955,Sandage1960,Sandage1966}. These results played an important role informing early stellar evolutionary models \citep{Hoyle1955,Kippenhahn1958,Demarque1963}.
It was thus established early on that GCs can provide important insights into both stellar and galactic astrophysics. 

As more data became available, the relatively low incidence of clusters with metallicities near $\mathrm{[Fe/H]}=-2.5$ became evident \citep{Bond1981}. By comparison with simple models for Galactic chemical evolution or with metallicity distributions of halo field stars, it has been estimated that there should be about a handful of GCs with $\mathrm{[Fe/H]}<-2.5$ in the Milky Way, while none is observed \citep{Carney1996,Simpson2018,Beasley2019,Youakim2020}. This has led to the notion of a ``metallicity floor'' for GCs near $\mathrm{[Fe/H]}\approx-2.5$. It has been suggested that such a  metallicity floor may be a consequence of the hierarchical nature of galaxy assembly, combined with the galaxy mass--metallicity relation \citep{Harris2006a,Choksi2018,Usher2018,Kruijssen2019}. In this scenario, dwarf (proto-) galaxies with metallicities less than $\mathrm{[Fe/H]}=-2.5$ that contributed to the build-up of galactic halos would be insufficiently massive to form massive GCs that could survive for a Hubble time. These ideas appeared to be supported by the identification of the Phoenix stream in the Milky Way halo as the possible remnant of a low-mass GC with $\mathrm{[Fe/H]}=-2.7$ \citep{Wan2020}.

From an integrated-light spectrum obtained with the HIRES spectrograph on the Keck~I telescope, the globular cluster EXT8 in Messier 31 (M31) was recently shown to have a metallicity of $\mathrm{[Fe/H]}=-2.91\pm0.04$, well below the metallicity floor \citep{Larsen2020}. Unlike the Phoenix stream progenitor, EXT8 is fairly massive, with an estimated dynamical mass of $(1.14\pm0.16)\times10^6 M_\odot$, and thus represents a challenge to the notion that metal-poor, massive GCs could not have formed in the early Universe.  The cluster was first mentioned in the literature as EX8, part of an ``external survey'' \citep{Federici1990}. It was designated EXT8 by \citet{Battistini1993} and lies at a projected distance of 27~kpc from the centre of M31. Based on an analysis of the spatial distribution and kinematics of GCs in the M31 halo, EXT8 does not appear to be associated with known M31 halo substructure \citep{Mackey2019a}.
While metallicities between $\mathrm{[Fe/H]}=-2.80$ and $\mathrm{[Fe/H]}=-2.07$ had previously been reported for EXT8  from spectroscopic analyses \citep{Fan2011,Chen2016}, the HIRES spectrum showed very clearly that the metallic lines are much weaker than in the metal-poor Galactic GC M15. 
Most of the $\alpha$-elements (Si, Ca, Ti) were found to be enhanced by about a factor of two compared to scaled-solar composition, as is typical for metal-poor old stellar populations, but Mg was found to be strongly deficient with $\mathrm{[Mg/Fe]}=-0.35\pm0.05$  \citep{Larsen2020}. Although the abundances of the $\alpha$-elements do not everywhere vary strictly in lockstep \citep{Villaume2020}, it is very unusual for Mg to be as discrepant as observed for EXT8, and a similar combination of low $\mathrm{[Mg/Fe]}$ and enhancement of other $\alpha$-elements is not observed in metal-poor Galactic field stars \citep{Frebel2015,Kobayashi2020}. One possibility is that the Mg deficiency is related to an extreme case of the Mg-Al anticorrelation and the ``multiple stellar populations'' phenomenon in GCs \citep{Bastian2018,Gratton2019}.

EXT8 provides us with a unique opportunity to explore the CMD of a stellar population with a metallicity near $\mathrm{[Fe/H]}=-3$ and compare with stellar model predictions at a metallicity that is about 0.5~dex lower than the current limit, set by the metallicity floor of Galactic GCs. The expectation is that the CMD should show a steep RGB and a blue HB, as is typical of metal-poor GCs. However, it has long been known that HB morphology is affected by parameters other than metallicity such as age and helium abundance \citep{Sandage1967,vandenBergh1967,Lee1994,Moehler2001,Catelan2009,Gratton2010} and there are examples of metal-poor GCs with relatively red HB morphologies, such as Fornax~1 in the Fornax dwarf spheroidal galaxy \citep{Buonanno1998,DAntona2013}. Interest in HB morphology as an indicator of helium abundance variations has been revived due to the realisation that it may be linked to the variations in the abundances of other light elements associated with the presence of multiple populations in GCs \citep{DAntona2002,DiCriscienzo2011,Gratton2010,Nardiello2019}.

In favourable cases, CMDs can be obtained for the outer parts of M31 GCs even from the ground \citep{Mackey2010a}, and {\it Hubble Space Telescope} ({\it HST}) observations can reach below the HB level in a few orbits of observing time \citep{Rich2005,Mackey2006}. One result from such work is that GCs in the outer halo of M31 tend to have redder HB morphologies at a given metallicity than their Galactic counterparts, possibly as a result of age differences \citep{Mackey2007a,Perina2012}.

Here we report on new observations of EXT8 that we have obtained with {\it HST}. Given the potential implications for understanding GC formation in the early Universe, our primary aim is to verify whether or not the CMD supports the spectroscopic evidence that EXT8 is an old, very metal-poor GC. A related aim is to put constraints on any metallicity spread that might be present in the cluster, as observed in some Milky Way GCs that are suspected to be the nuclei of disrupted dwarf galaxies and might thus have formed via a different channel than ``normal'' GCs \citep{Pfeffer2021}.  We also discuss the general appearance of the CMD and compare with predictions by various theoretical isochrones. 

\section{Observations}

\begin{figure}
\includegraphics[width=8.8cm]{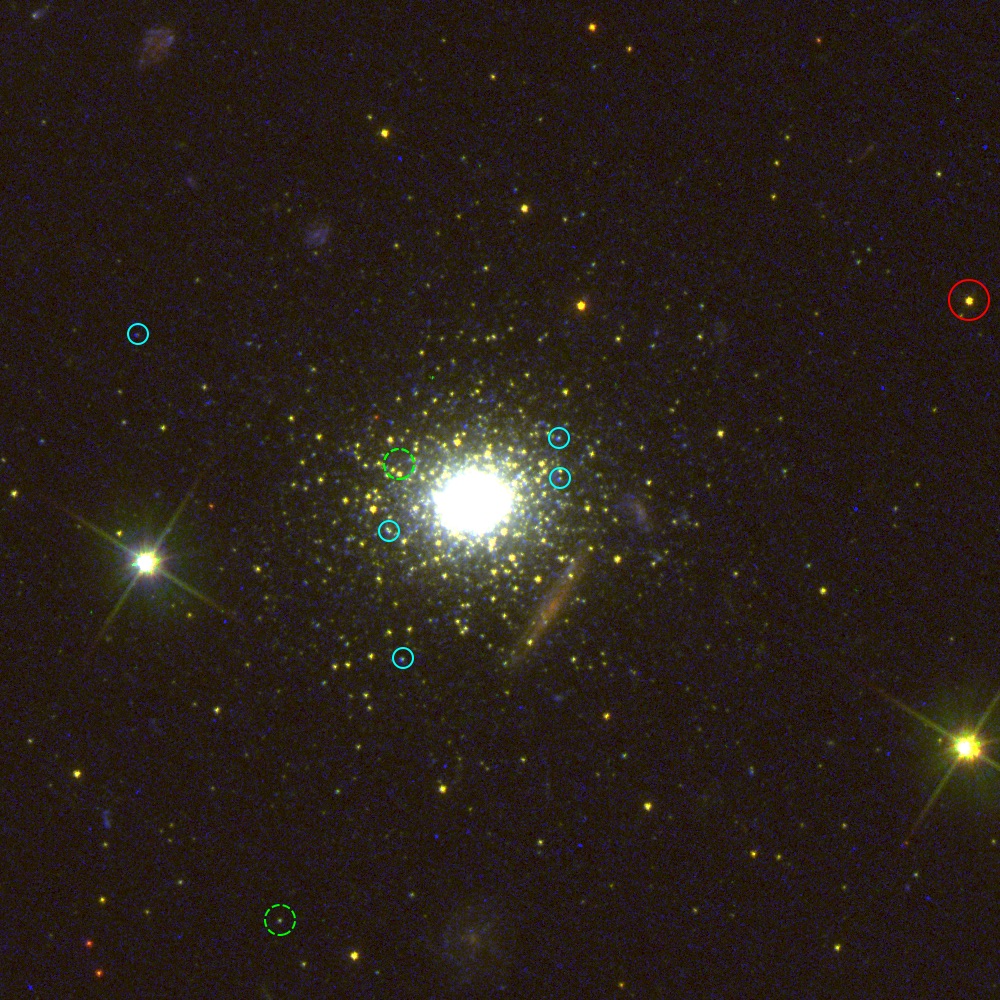}
\caption{\label{fig:img}\textcolor{blue}{}Colour image of EXT8 produced from the F300X, F606W, and F814W \textit{HST}/WFC3 images.
The figure shows a $40^{\prime\prime}\times40^{\prime\prime}$ subsection of the full WFC3 field of view, or about 152 pc$\times$152 pc at the assumed distance of M31. North is up and east to the left. The large red circle marks the brightest RGB star (Sect.~\ref{sec:trgb}), the small cyan circles mark the UV-bright stars discussed in Sect.~\ref{sec:hbuv}, and
the medium-sized (dashed) green circles mark two RR Lyrae candidates (Sect.~\ref{sec:rrlyr}).}
\end{figure}

EXT8 was observed with the Wide Field Camera 3 (WFC3) on board {\it HST} on 13-15 Jan 2021 (program ID 16459, PI S.\ S.\ Larsen). Exposures were obtained in the filters F300X, F606W, and F814W, using two orbits each for the F300X and F606W observations and three orbits for F814W. The observations were split into two sub-exposures per orbit for a total of four individual exposures in F300X and F606W and six in F814W, all of which were dithered according to the patterns in \citet{Anderson2020}.
The total exposure times were 5414~s (F300X), 5428~s (F606W), and 8128~s (F814W). EXT8 was placed on the UVIS2 detector and centred near detector coordinates $(x, y) = (1040, 1090)$, with an offset of about 40\arcsec\ from the centre of the detector towards read-out amplifier C to mitigate the effect of charge-transfer inefficiency. A FLASH=10 post-flash illumination was further added to the F300X exposures. 

The F606W and F814W observations were designed to allow photometry of stars on the upper part of the RGB, approximately down to the level of the HB, while the F300X observations were aimed primarily at detecting and characterising HB and other hot stars. While ultraviolet photometry of RGB stars has been employed extensively to characterise multiple stellar populations in Galactic and extra-galactic GCs \citep{Larsen2014a,Piotto2015,Niederhofer2017}, reaching the required precision at the distance of M31 would require much longer exposure times.

Figure~\ref{fig:img} shows a colour image of EXT8 based on the F300X, F606W, and F814W exposures, covering a $40\arcsec\times40\arcsec$ field of view centred on the cluster. It can be seen that EXT8 is well resolved into individual stars, although the cluster is quite compact and crowding becomes severe in the central few arcsec. The cluster is visibly flattened, as was already evident from ground-based imaging \citep{Larsen2020}. Apart from the two bright foreground stars, most stars in the image are members of EXT8 with minimal contamination from the general M31 halo. Several background galaxies are also visible in the vicinity of EXT8. 

\section{Analysis}

The observations were retrieved from the Mikulski Archive for Space Telescopes (MAST). For our analysis we used the standard pipeline processed images, corrected for charge transfer inefficiencies (the \texttt{\_drc} and \texttt{\_flc} images). 
Throughout the paper we assume a distance of 783~kpc for M31 \citep{Stanek1998} and a Galactic foreground extinction at the position of EXT8 (J2000.0 coordinates RA=$00^h53^m14.5^s$, Decl = $+41^\circ 33^\prime 24\farcs5$) of $A_\mathrm{F300X}=0.353$~mag, $A_\mathrm{F606W}=0.167$~mag, and $A_\mathrm{F814W}=0.103$ in the WFC3 filters \citep[][via the NASA/IPAC Extragalactic Database (NED)]{Schlafly2011}. One WFC3/UVIS pixel ($0\farcs040$) corresponds to a physical scale of 0.15~pc at the assumed distance.

\subsection{Resolved photometry -- F606W and F814W}
\label{sec:viphot}

\begin{figure}
\includegraphics[width=8.8cm]{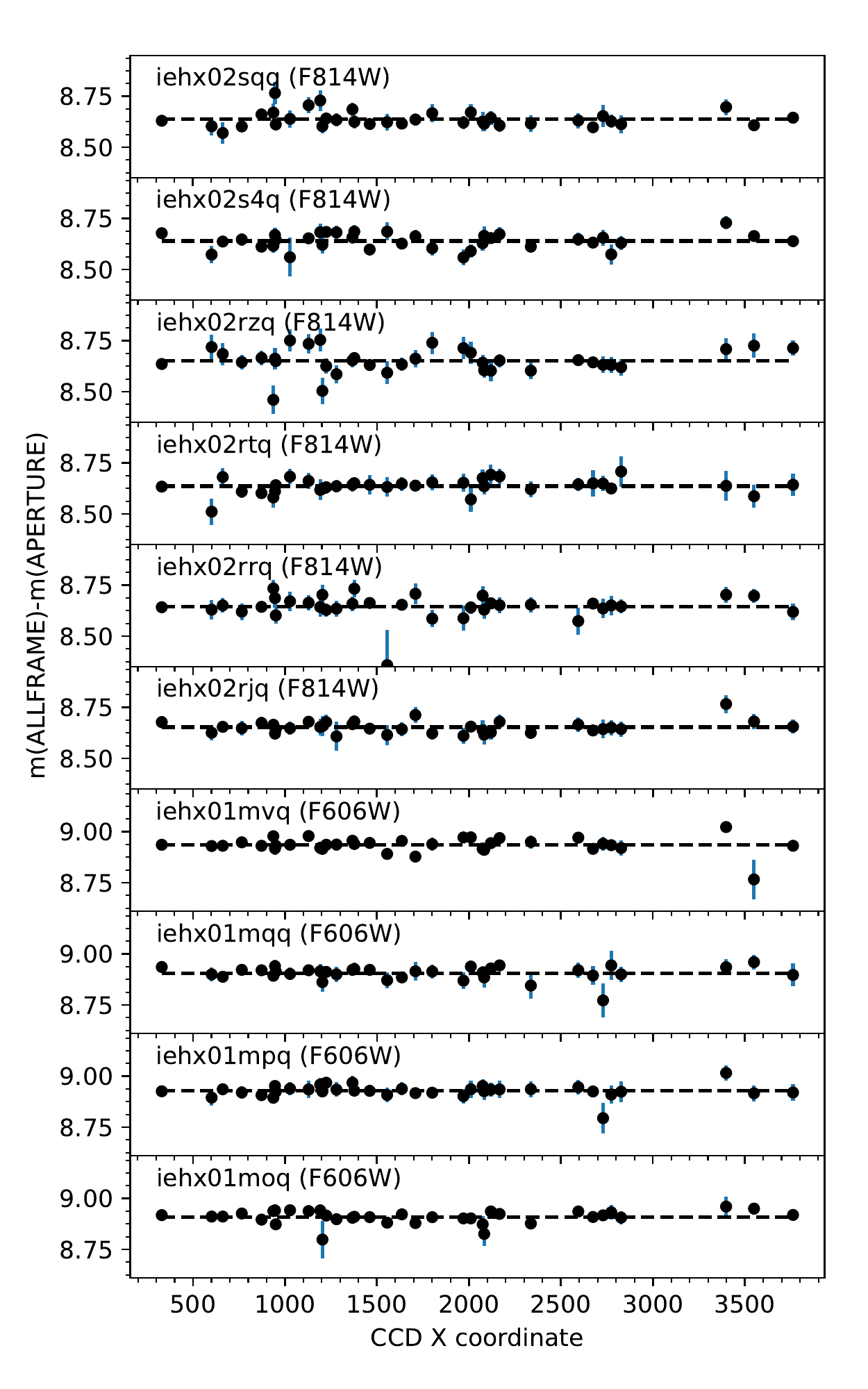}
\caption{\label{fig:cal}Zero-point corrections between \texttt{ALLFRAME} and aperture photometry for individual calibration stars in each frame. The ID of each exposure is given in the legend.
}
\end{figure}

We used \texttt{ALLFRAME} \citep{Stetson1994} to carry out point-spread function (PSF)-fitting photometry of stars in the individual F606W and F814W images (the \texttt{\_flc} files). EXT8 is fully contained within the UVIS2 detector and we did not carry out photometry on the UVIS1 detector. 
The procedure was similar to that adopted in \citet{Larsen2014a} and we refer to that paper for more details than those provided here. Briefly explained, the \texttt{\_flc} images were multiplied by the pixel area maps available from STScI \citep{Kalirai2010} and aligned to a common reference frame. 
As \texttt{ALLFRAME} solves for coordinate transformations between images, the main purpose of the alignment was to enable filtering of bad pixels and cosmic-ray events. Since sub-pixel shifts will tend to blur any sharp features, making them harder to remove, the images were only aligned to within integer pixels shifts. 
A master-frame was then produced for each filter by average combining the individual frames with the \texttt{imcombine} task in \texttt{IRAF} \citep{Tody1986}, using the \texttt{ccdclip} option to reject pixels that deviated by more than $5\sigma$ from the mean at each position. Next, pixels in the individual frames that were flagged as ``bad'' by \texttt{imcombine} were replaced by their values in the master frame. Stars were detected in the F606W master image with the \texttt{find} task in \texttt{DAOPHOT}, a PSF was produced with the \texttt{psf} task for each individual frame from 36 isolated, bright stars, and a first pass of \texttt{ALLFRAME} photometry was obtained with all F606W and F814W frames as input. Additional stars were detected in a second pass of \texttt{find} on the star subtracted F606W master frame, the PSFs were redetermined for each individual frame on star-subtracted images with only the PSF stars remaining, and a second pass of \texttt{ALLFRAME} photometry was obtained with the combined star list as input. The photometry was calibrated to the standard STMAG system by matching the PSF magnitudes to aperture photometry of the PSF stars in a 10-pixel radius ($0\farcs4$) aperture, applying photometric zero-points computed from the \texttt{PHOTFLAM} header keywords, and adding aperture corrections of $-0.102$ mag (F606W) and $-0.107$ mag (F814W) from the 10-pixel radius to infinity \citep{Hartig2009}. The star-to-star dispersion of the zero-point offsets between the aperture- and PSF-fitting photometry was about 0.03--0.04~mag per exposure (Fig.~\ref{fig:cal}), which is consistent with the uncertainties on the individual offsets. The corresponding  random uncertainties on the mean zero-point offsets are then less than 0.01~mag per exposure. The photometry from the individual frames was average combined to produce the final photometric catalogue, which is available at the CDS (Table~\ref{tab:photometry}). 

The photometric uncertainties and completeness were quantified by means of artificial star experiments. To this end, an artificial cluster was added to each \texttt{\_flc} image on the other half of the UVIS2 detector, opposite EXT8. To generate the artificial cluster, artificial stars were sampled at random from a \citet{King1962} profile with a half-light radius of $0\farcs91$ and a tidal radius of 15\arcsec, which implies a core radius of $0\farcs21$ \citep[Ishape User's Guide;][]{baolab}. These core- and half-light radii are similar to those found from 2-dimensional King profile fits to the images of EXT8 (Sect.~\ref{sec:iprop}). The tidal radius adopted for these experiments is, however, smaller than the outer radius of EXT8 (about 25\arcsec), as a King profile with a larger tidal radius produces too many stars in the outer parts of the cluster. This inconsistency is due to the fact that the actual cluster profile is not well fitted by a King profile in the outer parts (Sect.~\ref{sec:iprop}).
Each artificial star was assigned a mass drawn at random from a \citet{Kroupa2001} mass function, and F606W and F814W magnitudes were then assigned by interpolation in an $\alpha$-enhanced DSEP isochrone \citep[Dartmouth Stellar Evolution Program;][]{Dotter2007} with an age of 13~Gyr and $\mathrm{[Fe/H]}=-2.5$ (the lowest metallicity available). 
The total number of artificial stars was scaled such that the integrated magnitude of the artificial cluster was similar to that of EXT8. The DSEP isochrones do not include the HB, but we added a ``blue sequence'' of artificial stars at $m_\mathrm{F606W}-m_\mathrm{F814W} = -1.0$ and with magnitudes of $m_\mathrm{F606W}=25.5, 26.0, \ldots, 28.0$ to quantify the detection completeness and photometric uncertainties for stars on an extended HB. In each run, 5 such stars were included per magnitude step, restricting these stars to locations beyond a radius of 3\arcsec .
The artificial stars were then added to the UVIS2 images with the \texttt{mksynth} task in the \texttt{BAOLAB} package \citep{baolab} using PSFs obtained with \texttt{DAOPHOT}. A similar number of artificial PSF stars as those used for the photometry of EXT8 were also added to the images. Photometry was then carried out with \texttt{ALLFRAME} in the exact same way as for the analysis of EXT8. 
To improve statistics, the procedure was repeated 20 times with different random realisations of the artificial cluster and blue sequence, yielding a total of 100 stars for each magnitude step on the blue sequence. 

\subsection{Resolved photometry -- F300X}

As noted already by \citet{Sandage1955}, the $U$-band brightness of the HB is comparable to that of the RGB. Comparison with BaSTI isochrones \citep{Hidalgo2018,Pietrinferni2021} confirms that the brightness of HB stars is expected to remain fairly constant at $M_\mathrm{F300X}\approx +0.8$ over a range of colours, which is similar to the tip of the RGB in this band. 
Using the F300X images for source detection is therefore a very efficient way to obtain a fairly clean sample of HB stars, and the entire F300X image in fact contains few stars other than HB stars in EXT8. However, this also introduced some challenges for the photometry by making it difficult to align the individual images and find suitable PSF stars. Moreover, even the HB stars are relatively faint in the F300X images, and requiring a measurement in each individual F300X \texttt{\_flc} image would have further reduced the completeness. The photometry involving the F300X images was therefore done separately, following a somewhat different procedure than that outlined above. 

For the photometry involving the F300X data we carried out photometry directly on the drizzle-combined images (the \texttt{\_drc} files produced by the pipeline).
Even so, we could identify only six suitable PSF stars in the F300X image, forcing the assumption of a constant PSF across the UVIS2 detector. It is therefore inevitable that the F300X photometry is of lower quality than that obtained from the F606W and F814W data. The PSFs for the drizzled F606W and F814W images were determined separately using a larger number of PSF stars, as for the reduction based on the \texttt{\_flc} images, and all three \texttt{\_drc} images were fed to \texttt{ALLFRAME} and analysed simultaneously. The photometry on the drizzled images is available in Table~\ref{tab:uviphotometry}. 

A separate set of completeness tests were performed for the photometry involving the F300X image, mostly following the procedure outlined above for the F606W/F814W photometry. 
As before, a sequence of blue stars with $m_\mathrm{F606W}-m_\mathrm{F814W} = -1.0$ and $m_\mathrm{F606W}$ between 25.5 and 28.0 were added to the images. However, the detection completeness here depends much more strongly on the brightness in the F300X images and to quantify this we therefore also repeated the completeness tests for a range of F300X magnitudes,    $m_\mathrm{F300X}=25.0, 25.5, \ldots, 27.0$.

\subsection{Colour-magnitude diagrams for EXT8 and the artificial cluster}
\label{sec:cmd0}

\begin{figure}
\includegraphics[width=89mm]{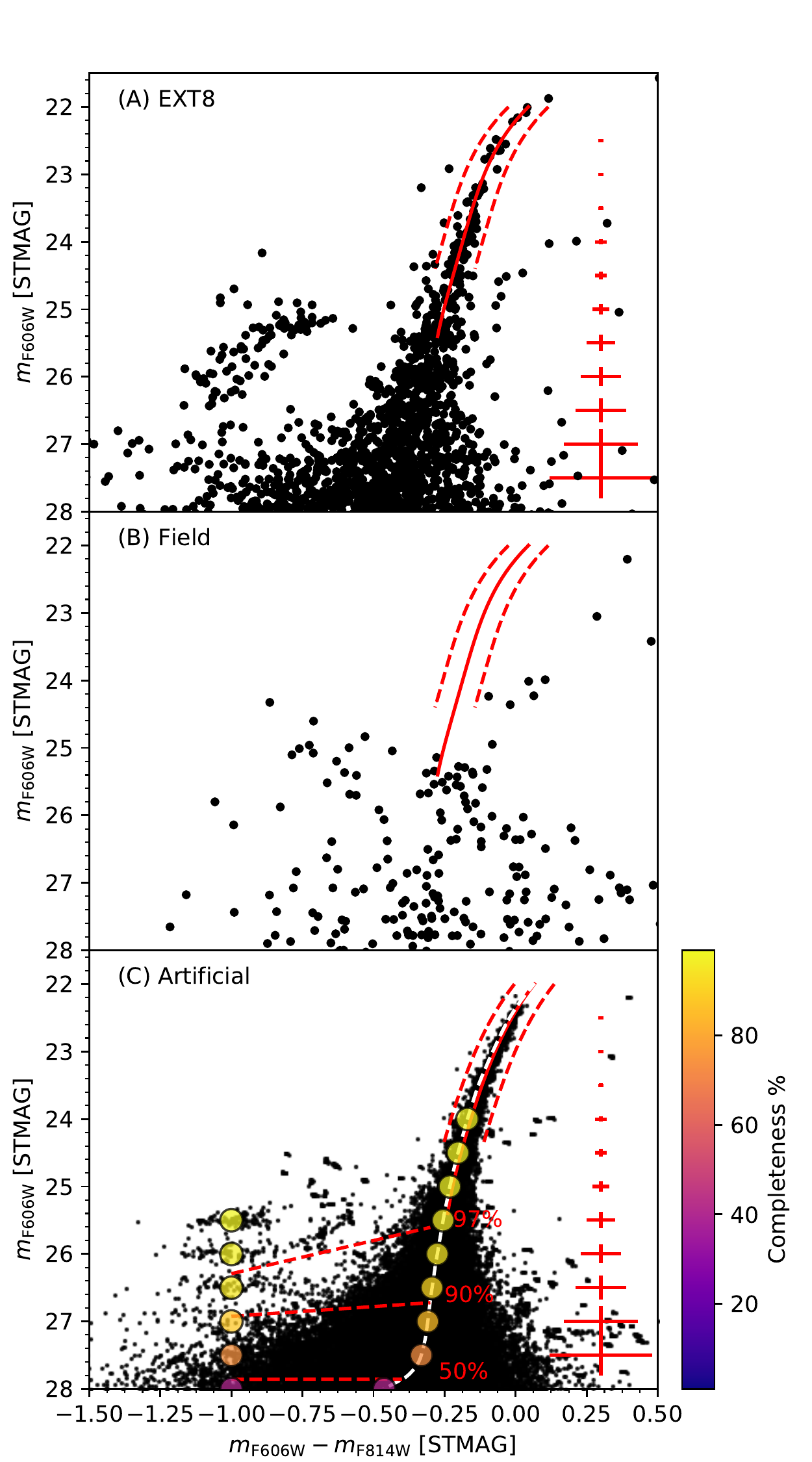}
\caption{\label{fig:cmds}Colour--magnitude diagrams for EXT8 (A), a field region (B), and the artificial cluster used to quantify completeness and photometric errors (C). Ridge lines (polynomial fits) are shown with red curves and the parallel dashed curves mark the regions within which colour spreads are shown in Fig.~\ref{fig:dvihist}. 
In panel C, the coloured circles indicate the completeness according to the scale on the right. The 50\%, 90\%, and 97\% completeness levels on the blue sequence and on the isochrone are connected with dashed lines. Error bars indicate the dispersion of the measured colours and magnitudes for the artificial stars around their input values.}
\end{figure}

Figure~\ref{fig:cmds} shows the colour--magnitude diagrams for the F606W/F814W photometry of EXT8 (panel A), a field region centred on the position at which the artificial cluster was added in the artificial star experiments (panel B), and the combined CMD for the 20 artificial cluster realisations (panel C).
In this figure, and throughout the remainder of this paper, photometry is shown for distances in the range $3\arcsec < r < 24\arcsec$ from the centre of EXT8 (or the artificial cluster). 
At radii less than 3\arcsec, the stellar density is too high to obtain reliable photometry and at radii greater than about 25\arcsec\ very few cluster members remain. The RGB and HB of EXT8 are clearly visible in panel A and will be discussed in more detail below (Sect.~\ref{sec:cmds}). 

The solid red curve in the top and centre panels is a ridge line for the upper EXT8 RGB with dashed lines offset by $\pm0.07$~mag in $m_\mathrm{F606W}-m_\mathrm{F814W}$. The ridge line is a fourth-order polynomial fit to the colours of the RGB stars brighter than $m_\mathrm{F606W}=25.5$. To define the ridge line we used an iterative procedure to reject stars that deviated by more than 0.1~mag from the polynomial fit. 
Since all panels in Fig.~\ref{fig:cmds} cover the same area on the sky, it can be seen from a direct comparison of panels A and B that contamination from the general M31 halo field is expected to be very minimal for stars on the upper RGB of EXT8, while the redder part of the HB may be contaminated by a few non-members. A few stars located on the red side of the RGB in panel A also have counterparts in panel B and are likely field stars. 

The ridge line in panel C was obtained in the same way as that in panel A, but here from a fit to the artificial star photometry. It coincides closely with the isochrone used to generate the input photometry for the artificial star experiments (white dashed curve).  The coloured circles indicate the detection completeness determined for the sequence at $m_\mathrm{F606W}-m_\mathrm{F814W} = -1.0$ and for stars along the isochrone in 0.5~mag bins of $m_\mathrm{F606W}$. The 50\%, 90\%, and 97\% completeness levels determined for stars on the blue sequence and the isochrone are connected with dashed lines.
For both the blue sequence and the isochrone, the completeness at the faint limit of the plot range, $m_\mathrm{F606W}=28$, is about 38\%. The 90\% completeness limit is at $m_\mathrm{F606W}\approx26.8$ and for RGB stars brighter than $m_\mathrm{F606W}=25.5$ the completeness is greater than 97\%. As these completeness fractions were determined from the artificial cluster, they represent an overall level of completeness for all stars with $r > 3\arcsec$.

In addition to the artificial stars, the real stars from panel B are again visible in panel C. Due to small variations in the PSF reconstruction and the photometric zero-point calibrations in the 20 realisations of the artificial star photometry, the magnitudes and colours of these stars differ slightly between the 20 realisations. The dispersions of the magnitudes and colours are less than 0.005~mag, which provides an additional estimate of the internal precision of the photometric calibration. However, the total uncertainty is likely somewhat larger. 

\subsection{Photometric uncertainties}
\label{sec:artex}

\begin{figure}
\includegraphics[width=89mm]{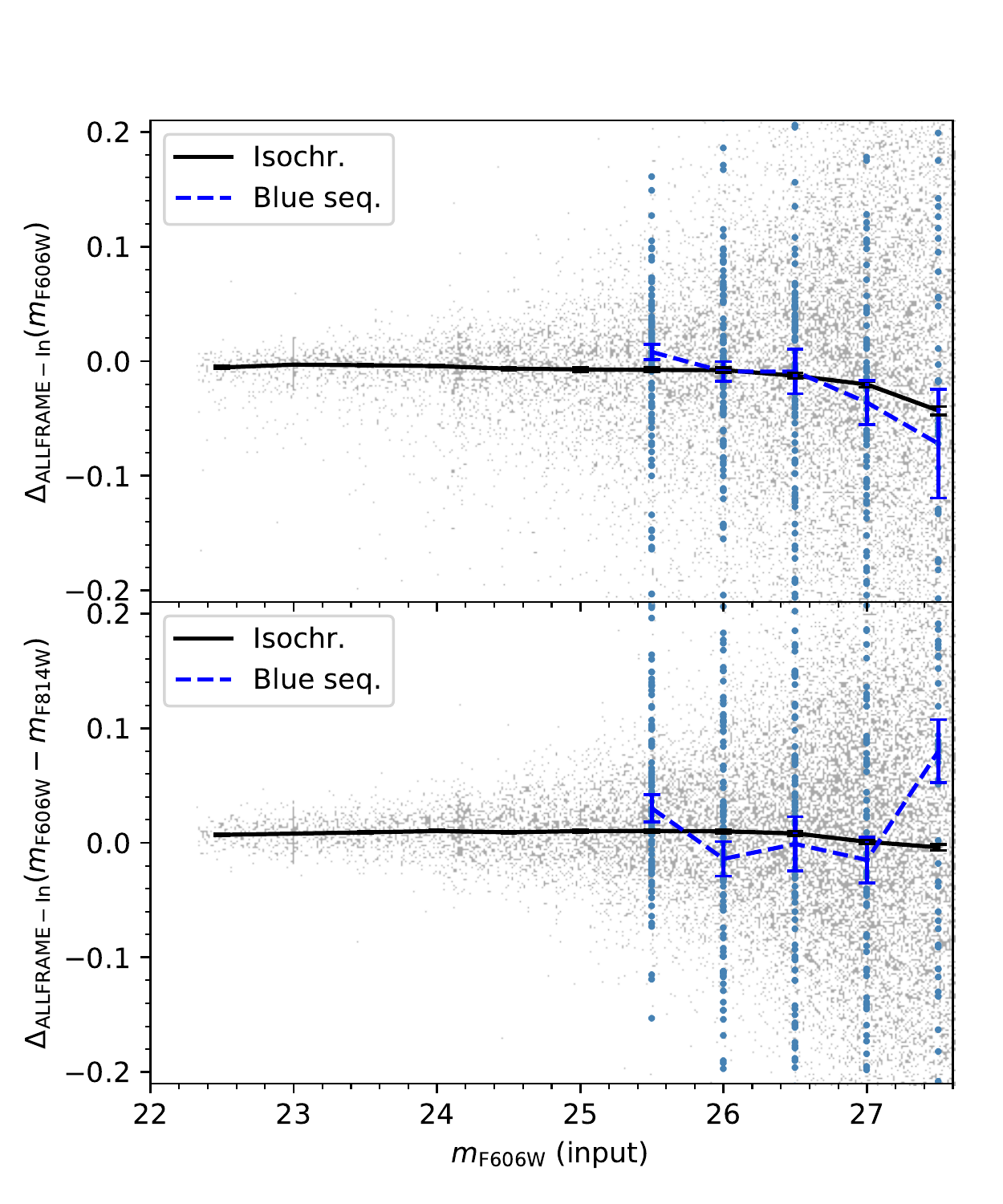}
\caption{\label{fig:v_dvi}Difference between input and \texttt{ALLFRAME} measurements of $m_\mathrm{F606W}$ magnitudes (top) and $m_\mathrm{F606W}-m_\mathrm{F814W}$ colours (bottom)
for artificial star experiments.
Solid black lines show median values for artificial stars sampled from isochrones (grey points) while dashed blue lines show median values for artificial stars on the blue sequence (blue dots).}
\end{figure}

\begin{figure}
\includegraphics[width=89mm]{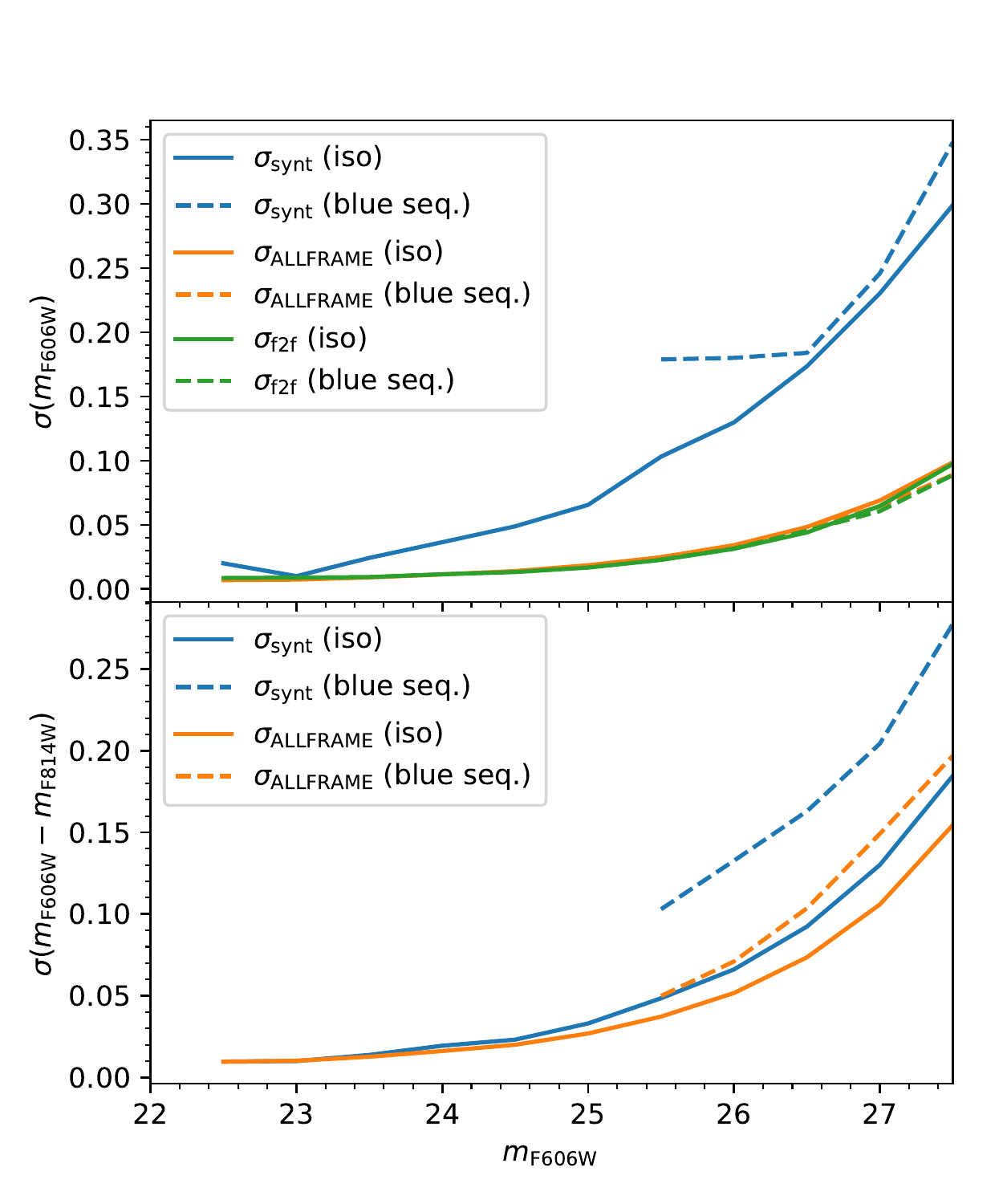}
\caption{\label{fig:errcmp}Comparison of error estimates for artificial star tests: $\sigma_\mathrm{synt}$ are the dispersions of the \texttt{ALLFRAME}-Input magnitudes, $\sigma_\mathrm{ALLFRAME}$ are the \texttt{ALLFRAME} errors propagated to the mean magnitudes, and $\sigma_\mathrm{f2f}$ are the errors on the mean \texttt{ALLFRAME} magnitudes as determined from the frame-to-frame dispersion. Solid lines are for stars sampled from isochrone while dashed lines are for stars on blue sequence.}
\end{figure}

In Fig.~\ref{fig:v_dvi} we compare the input colours and magnitudes for the artificial cluster stars with the \texttt{ALLFRAME} measurements. The small grey points show the differences between measured and input values, $\Delta_\mathrm{ALLFRAME-In}$, for stars sampled from the isochrone, and larger light blue dots are for stars on the blue sequence. The solid black and blue lines show the median differences in bins of 0.5~mag with error bars determined from 100 Monte-Carlo experiments. For stars on the isochrone, the error bars on the median values are generally smaller than the width of the black line. 
The sharp feature at $m_\mathrm{F606W}=23$ is due to the artificial PSF stars while the fainter feature near $m_\mathrm{F606W}\approx24.2$ is a real feature of the RGB luminosity function, the RGB ``bump'' \citep{Bjork2006}.
The distribution of $\Delta_\mathrm{ALLFRAME-In}(m_\mathrm{F606W})$ values is somewhat asymmetric with a larger scatter towards negative offsets (brighter \texttt{ALLFRAME} magnitudes), such as would be caused by unresolved blends, and we have therefore  chosen to use the median values as they are less sensitive to outliers.

Systematic trends in the $\Delta_\mathrm{ALLFRAME-In}$ values versus input magnitude are generally very minor. For stars brighter than the HB level, $m_\mathrm{F606W} \la 25$, the median magnitude offsets are between $\mathrm{med}\left[ \Delta_\mathrm{ALLFRAME-In}(m_\mathrm{F606W})\right] = (-0.003\pm0.001)$~mag  and ($-0.007\pm0.001$)~mag (top panel) and the colour offsets  are between 
$\mathrm{med}\left[\Delta_\mathrm{ALLFRAME-In}(m_\mathrm{F606W}-m_\mathrm{F814W})\right]$ = $(+0.007\pm0.001)$~mag and $+0.010\pm0.001$~mag (bottom panel).
The zero-points of the input magnitudes are somewhat poorly defined because the empirical PSFs used to generate the artificial stars are imperfect approximations to the true PSF. This likely accounts for any small overall differences in the zero-points of the input- and measured magnitudes. The median offsets remain small for stars sampled from the isochrones also at fainter magnitudes, reaching $-0.04$~mag in $\Delta_\mathrm{ALLFRAME-In}(m_\mathrm{F606W})$ and $-0.004$~mag in $\Delta_\mathrm{ALLFRAME-In}(m_\mathrm{F606W}-m_\mathrm{F814W})$ at $m_\mathrm{F606W}=27.5$. For the blue sequence the uncertainties are larger because of the smaller numbers of artificial stars, but the median $\Delta_\mathrm{ALLFRAME-In}$ values are mostly consistent with those seen for the stars sampled from the isochrones. 

In Fig.~\ref{fig:errcmp} we compare three different ways of estimating the photometric uncertainties: $\sigma_\mathrm{ALLFRAME}$ was obtained by propagating the errors on the individual measurements reported by \texttt{ALLFRAME}, $\sigma_\mathrm{f2f}$ was computed from the frame-to-frame dispersions of the magnitude measurements for each star, and $\sigma_\mathrm{synt}$ is the dispersion of the $\Delta_\mathrm{ALLFRAME-In}$ values. In general, the $\sigma_\mathrm{ALLFRAME}$ and $\sigma_\mathrm{f2f}$ estimates are very similar, which shows that the errors reported by \texttt{ALLFRAME} accurately represent the \emph{random} uncertainties on the photometry. However, both are significantly smaller than $\sigma_\mathrm{synt}$. The difference is most pronounced for the $m_\mathrm{F606W}$ magnitude measurements, where $\sigma_\mathrm{synt}$ is 2--3 times greater than $\sigma_\mathrm{ALLFRAME}$ and $\sigma_\mathrm{f2f}$ (top panel). For the $m_\mathrm{F606W}-m_\mathrm{F814W}$ colours the agreement is better, but the $\sigma_\mathrm{synt}$ estimates remain somewhat larger than the \texttt{ALLFRAME} errors also here. These differences between the errors based on the photometry alone and those based on the comparison with artificial star tests are most likely due to the effect of unresolved blends and crowding, which can introduce a bias on the magnitude measurements without necessarily affecting the random uncertainties strongly. Indeed, we found that the $\sigma_\mathrm{synt}$ values agreed more closely with $\sigma_\mathrm{ALLFRAME}$ and $\sigma_\mathrm{f2f}$ for stars further from the centre of the artificial cluster. For example, at radii $>8\arcsec$ the $\sigma_\mathrm{synt}$ errors on the $m_\mathrm{F606W}$ magnitudes exceed $\sigma_\mathrm{ALLFRAME}$ by only $\sim$50\%. We note, also, that all three uncertainty estimates agree well for the bin at $m_\mathrm{F606W}=23$.  This bin contains the artificial PSF stars, which were deliberately placed in uncrowded parts of the image. 
We further note that the uncertainties on the $m_\mathrm{F606W}$ magnitudes are similar for stars sampled from the isochrone and from the blue sequence, whereas the uncertainties on the colours are greater for stars on the blue sequence. The latter is explained by the fact that stars on the blue sequence are fainter in the F814W images.

In summary, the \texttt{ALLFRAME} photometry reproduces the input colours of the artificial stars without significant biases ($<0.01$~mag) across the relevant parts of the CMD. The full uncertainties are, however, larger than the formal errors computed by \texttt{ALLFRAME} and are therefore better quantified using the artificial star experiments.

\section{Results}

\begin{figure*}
\centering
\includegraphics[width=18cm]{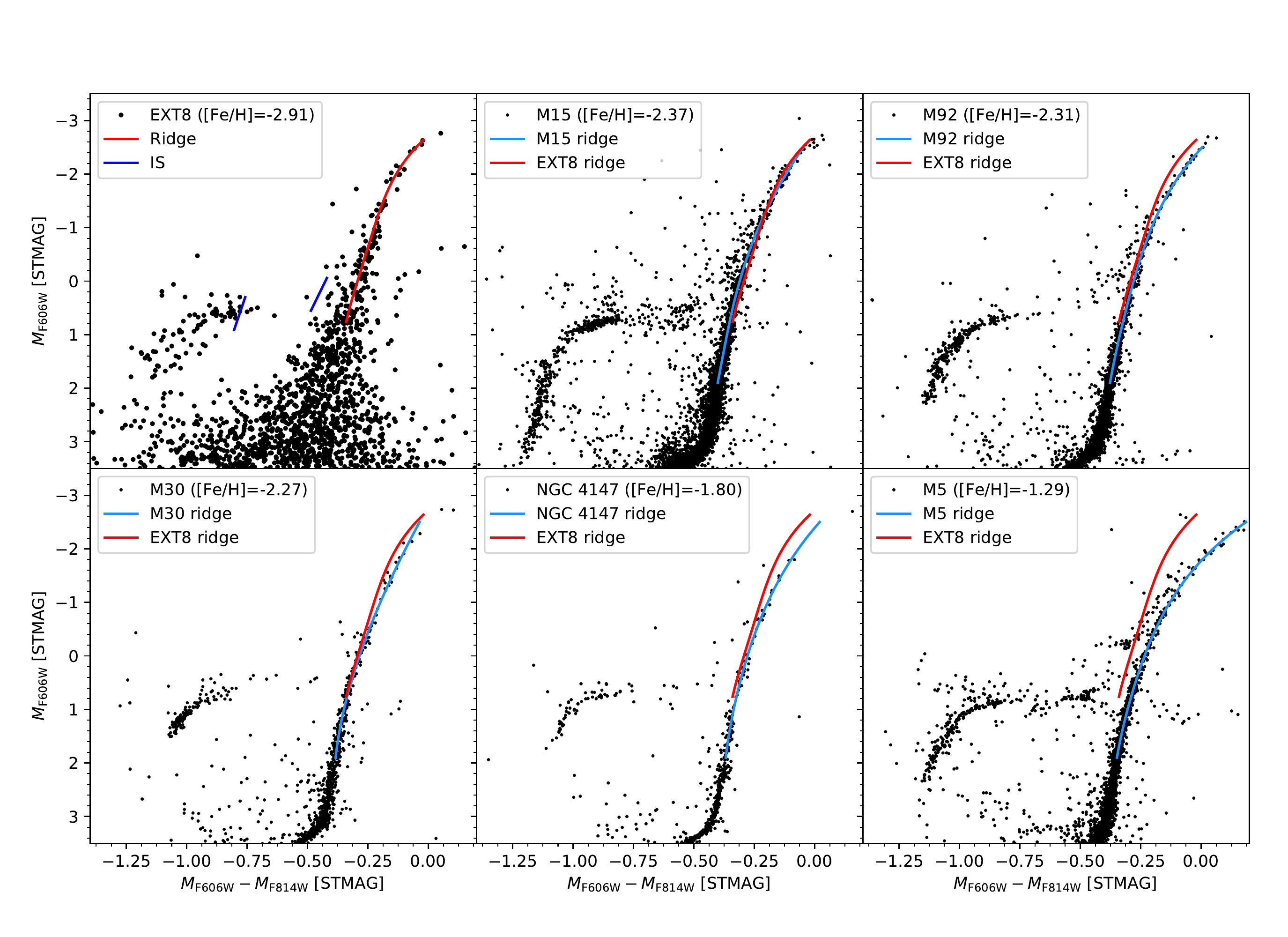}
\caption{\label{fig:cmdcmp}Colour--magnitude diagrams of EXT8 and the Galactic GCs M15, M92, M30, NGC~4147, and M5 (in order of increasing metallicity). In each panel, the red curve is the EXT8 ridge line and the blue curves are the ridge lines for the Galactic GCs.  The boundaries of the instability strip (IS) are indicated in the CMD of EXT8.}
\end{figure*}

\label{sec:cmds}

\begin{table}
\caption{Data for Galactic globular clusters.}
\label{tab:gcdata}
\centering
\begin{tabular}{lclll}
\hline\hline
 Cluster & Distance$^a$ & $A_\mathrm{F606W}^b$ & $A_\mathrm{F814W}^b$ & $\mathrm{[Fe/H]}^{c}$\\
 & (kpc) & (mag) & (mag) \\
\hline
M5 & 7.3 & 0.091 & 0.056 & $-1.29$ \\
M15 & 10.3 & 0.270 & 0.167 & $-2.37$ \\
M30 & 8.1 & 0.126 & 0.078 & $-2.27$ \\
M92 & 8.3 & 0.055 & 0.034 & $-2.31$\\
NGC 4147 & 19.3 & 0.064 & 0.040 & $-1.80$\\
\hline
\end{tabular}
\tablefoot{$^a$: Distances from \citet[][2010 revision]{Harris1996}, except M15 \citep{vandenBosch2006}. $^b$: \citet{Schlafly2011} via NED. $^c$: \citet{Harris1996}.
}
\end{table}

In Fig.~\ref{fig:cmdcmp} we show the CMD of EXT8 together with those of the Galactic GCs M15, M92, M30, NGC~4147, and M5 (in order of increasing metallicity).  These clusters have been selected to have relatively small foreground extinctions, so as to reduce uncertainties in the comparison of the CMDs. Their distances, foreground extinctions, and metallicities are listed in Table~\ref{tab:gcdata}.
M15, M30, and M92 are among the most metal-poor GCs in the Milky Way while 
NGC~4147 and M5 are about 0.5~dex and 1.0~dex more metal-rich, respectively. 
The composition of NGC~4147 has been less extensively studied than those of the other clusters, but a  recent determination of its metallicity ($\mathrm{[Fe/H]}=-1.84\pm0.02$; \citealt{Villanova2016}) agrees well with the value in the \citet{Harris1996} catalogue given in the table.
The photometry for the Galactic GCs was transformed from the catalogues of F606W and F814W magnitudes provided by the ACS Galactic Globular Clusters Survey \citep[ACSGCS;][]{Sarajedini2007, Anderson2008} to the WFC3 equivalents using the relations in \citet{Deustua2018}.
The transformed VEGAMAG magnitudes from the ACSGCS data were further converted to the STMAG system by adding the differences between the corresponding zero-points ($\Delta_\mathrm{F606W} = 0.246$~mag and $\Delta_\mathrm{F814W} = 1.259$~mag), taken from the WFC3/UVIS web pages at STScI\footnote{\url{https://www.stsci.edu/hst/instrumentation/wfc3/data-analysis/photometric-calibration/uvis-photometric-calibration}}.

For each Galactic GC we show an RGB ridge line together with the  ridge line for EXT8, again obtained as polynomial fits. We discuss the RGBs in more detail below (Sect.~\ref{sec:rgbs}) but it is already clear from Fig.~\ref{fig:cmdcmp} that the differences below $\mathrm{[Fe/H]}=-2$ are fairly minor.
The EXT8 RGB is very similar in colour to that of M15, and both are slightly bluer than the M92 and M30 RGBs. For NGC~4147 and, especially, for M5, the RGBs are noticeably redder, as expected for their higher metallicities. Visually, the EXT8 ridge line appears slightly steeper than for the other GCs.

The HB will be discussed in Sect.~\ref{sec:hb}. To first order, the HB morphology of EXT8 resembles those of M30, M92, and NGC~4147. The detected HB stars lie mostly on the blue side of the instability strip (IS), the edges of which are marked on the EXT8 CMD in Fig.~\ref{fig:cmdcmp}.
A few stars are located within the IS; these will be considered further in Sect.~\ref{sec:rrlyr}. There is no evident excess of HB stars on the red side of the IS, although it cannot be excluded that a few red HB stars are present.
The edges of the IS were calculated using the relations in \citet{Marconi2015}, with some extrapolation to the lower metallicity. The blue and red edges are at $T_\mathrm{eff} \approx 7475$~K and $T_\mathrm{eff} \approx 5888$~K, which corresponds to $M_\mathrm{F606W}-M_\mathrm{F814W} = -0.8$ and $M_\mathrm{F606W}-M_\mathrm{F814W} = -0.5$ from interpolation in the colour--$T_\mathrm{eff}$ relation for stars on the HB of an $\mathrm{[Fe/H]}=-3.2$ BaSTI isochrone. It is not obvious from Fig.~\ref{fig:cmdcmp} whether the HB of EXT8 has an extended blue part, as in M15, and we will address this point in more detail based on the F300X data in Sect.~\ref{sec:hbuv}. 

The dynamically determined mass-to-light ratio and the hydrogen Balmer lines already strongly imply that EXT8 is an old, metal-poor GC \citep{Fan2011,Chen2016,Larsen2020}, and this is further confirmed by the general morphology of its CMD.

\subsection{The red giant branch}
\label{sec:rgbs}

\subsubsection{Comparison with Milky Way GCs}
\label{sec:rgbemp}

\begin{table}
\caption{RGB colour offsets with respect to M92 ridge line.}
\label{tab:rgbcol}
{\small
\centering
\begin{tabular}{lcc}
\hline\hline
 Cluster & \multicolumn{2}{c}{$\Delta \langle M_\mathrm{F606W}-M_\mathrm{F814W}\rangle_\mathrm{M92}$} \\
 & $-2.5<M_\mathrm{F606W} < -1.0$ & $+1.0<M_\mathrm{F606W} < +1.5$ \\
\hline
EXT8 & $-0.038\pm0.004$ & $-0.030\pm0.006$ \\
M15 & $-0.037\pm0.002$ & $-0.028\pm0.002$ \\
M92 & $-0.006\pm0.002$ & $-0.000\pm0.002$ \\
M30 & $-0.003\pm0.003$ & $-0.009\pm0.003$ \\
NGC 4147 & $+0.013\pm0.003$ & $-0.001\pm0.002$ \\
M5 & $+0.093\pm0.005$ & $+0.027\pm0.002$\\
\hline
\end{tabular}
}
\end{table}

To quantify the differences between the RGB colours of the various clusters we first measured the mean colour offsets with respect to the M92 RGB ridge line, $\Delta \langle M_\mathrm{F606W}-M_\mathrm{F814W}\rangle_\mathrm{M92}$, in two bins of absolute magnitude (Table~\ref{tab:rgbcol}). 
The brighter bin ($-2.5<M_\mathrm{F606W}<-1.0$) mostly lies above luminosities where the asymptotic giant branch (AGB) may bias the measurements of RGB colours (see, for example, the CMDs for M15 and M92), while the fainter bin ($+1.0 < M_\mathrm{F606W} < +1.5$) lies below the HB. 
EXT8 does indeed have the bluest RGB of the six clusters, but only barely: M15 has  colour offsets nearly identical to those of EXT8. The RGBs of both clusters are about 0.03~mag bluer than those of M30 and M92.
We note that the brighter bin has a small offset even for M92 itself, which illustrates the uncertainties related to the different ways of quantifying the RGB colours (polynomial fits vs.\ average colours of stars in some magnitude range). 

Given the low metallicity of EXT8, it is not surprising that it should have a very blue RGB.
The reason why the RGB of M15 matches EXT8 more closely than those of M30 and M92 is less clear, given the similar metallicities of M15, M30, and M92. From the same ACSGCS photometry used here, \citet{Vandenberg2013} found that M15 also had slightly bluer main sequence turn-off (MSTO) colours compared to M30 and M92. They pointed out that a decrease in the reddening correction towards M15 of about 0.02 mag in $E(B-V)$ (also corresponding to a difference in $E$(F606W$-$F814W)$\approx0.02$) would bring the MSTO colours of the three clusters into closer agreement. Given the higher foreground reddening towards M15, it may be reasonable to suspect that the uncertainty on its reddening correction is somewhat larger. A small redwards shift in the colours of M15 would then make EXT8 bluer than all three metal-poor Galactic GCs.

One might wonder how robust the RGB colour offsets measured for EXT8 with respect to the Galactic GCs are.
\citet{Anderson2008} found that ``unmodellable'' PSF variations due to focus changes, geometric distortions, and detector inhomogeneities can introduce zero-point uncertainties of up to 0.02~mag across the field-of-view for the ACSGCS photometry of Galactic GCs. 
The WFC3 is not immune to such effects \citep{Sabbi2013}, and while \citet{Anderson2008} had access to large numbers of bright PSF stars across the field of view, we are forced to rely on fainter stars, so that our PSFs are necessarily of lower fidelity. 
While the scatter in the photometric zero-points among the calibration stars is largely consistent with the individual 0.03--0.04~mag uncertainties (Sect.~\ref{sec:artex}), this could hide systematic effects at the level of $\sim0.01$~mag (Fig.~\ref{fig:cal}).
It therefore cannot be excluded that zero-point uncertainties of 0.01--0.02 mag are present in our photometry of EXT8. 
If the reddening of EXT8 is lower than we have assumed, this could in principle also produce a colour shift. However, a shift of 0.03--0.04~mag relative to the value of $E(\mathrm{F606W}-\mathrm{F814W}) = 0.064$ adopted here would require the reddening to be overestimated by about a factor of two, which seems unlikely. Indeed, the \citet{Schlafly2011} reddening at the position of EXT8 is $E(B-V) = 0.06$~mag, which is similar to or slightly \emph{less} than the reddenings found for other GCs around M31 from CMD analyses \citep{Mackey2006,Mackey2007a}. 
Differences between the ACS and WFC3 systems are probably only a minor contributor to the uncertainties. The ACS and WFC3 F814W systems are essentially identical \citep{Deustua2018}, while the conversion between the ACS and WFC3 F606W systems involves a slight colour dependence of up to 0.02~mag for a VEGAMAG colour of $m_\mathrm{F606W,VEGA}-m_\mathrm{F814W,VEGA}=1$, typical of RGB stars. After accounting for this, as we also do here, \citet{Deustua2018} found consistency between the photometric calibrations of ACS and WFC3 at the 0.5\% level, corresponding to an uncertainty of 0.005~mag. We also recall (Fig.~\ref{fig:v_dvi}) that systematic biases in the \texttt{ALLFRAME} photometry as a function of magnitude are below the 0.01~mag level for RGB stars.
In summary, then, it seems unlikely that the colour difference between the EXT8 RGB and the metal-poor Galactic GCs can be fully explained by uncertainties in the measurements, so that a real difference is likely present. 

Differences in metallicity are expected to be accompanied not only by a shift in mean RGB colours but also by a change in the slope of the RGB \citep{Sandage1960,Demarque1963,Hartwick1968,Saviane2000,Streich2014}. 
From a practical point of view, the slope has the advantage of being much less sensitive to zero-point uncertainties than the absolute colours are. For EXT8, the difference in colour offset between the bright and faint bins is $\Delta_\mathrm{F606W-F814W}\mathrm{(bright-faint)} = -0.008\pm0.007$~mag, while the corresponding values for the Galactic GCs are, in order of increasing metallicity: 
$-0.009\pm0.003$~mag (M15), 
$-0.006\pm0.003$~mag (M92), 
$+0.006\pm0.004$~mag (M30), 
$+0.014\pm0.004$~mag (NGC~4147) and 
$+0.066\pm0.005$~mag (M5). 
This is consistent with a gradual decrease in RGB slope with increasing metallicity,
although the differences between the most metal-poor GCs are not highly significant.

\begin{figure}
\includegraphics[width=89mm]{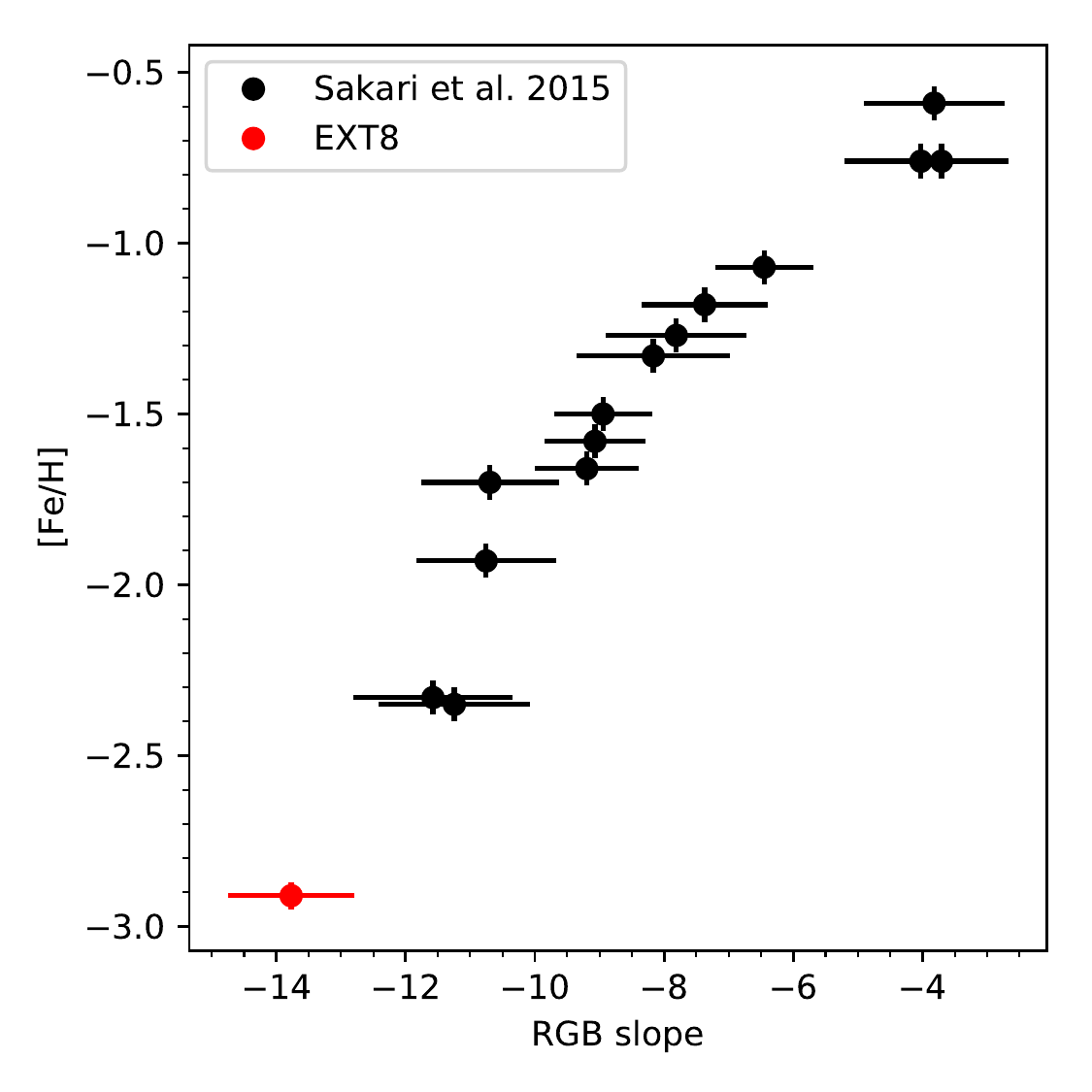}
\caption{\label{fig:fehslope}Metallicity versus red giant branch slope for Galactic GCs \citep{Sakari2015} and EXT8.}
\end{figure}

The RGB slopes for a larger sample of Milky Way and M31 GCs were quantified by \citet{Sakari2015}, using the ACSGCS photometry. They defined the slopes by measuring the $M_\mathrm{F606W} - M_\mathrm{F814W}$ RGB colours at two absolute magnitudes, $M_\mathrm{F606W,VEGA}=-2$ and $M_\mathrm{F606W,VEGA}=0$. Two of their GCs, M15 and M92, are in common with those studied here, and we have verified that we get similar slopes to those measured by \citet{Sakari2015} for these clusters.
For the EXT8 RGB ridge line the colours at the reference points are $M_\mathrm{F606W}-M_\mathrm{F814W} = -0.168$ and $-0.313$, and the resulting RGB slope is $\Delta (M_\mathrm{F606W})/\Delta (M_\mathrm{F606W}-M_\mathrm{F814W}) = -13.8\pm0.9$.
In Figure~\ref{fig:fehslope} the metallicities of EXT8 and the Milky Way GCs are plotted as a function of the RGB slopes. The slopes from \citet{Sakari2015} have been corrected for the $\sim2\%$ difference between the ACS and WFC3 systems. Measured in this way, the difference in RGB slope between EXT8 and the Milky Way GCs appears more significant and the EXT8 RGB slope is consistent with an extension of the relation defined by the Milky Way GCs to the lower metallicity.

\subsubsection{Comparison with theoretical isochrones}
\label{sec:isocmp}

\begin{figure}
\includegraphics[width=89mm]{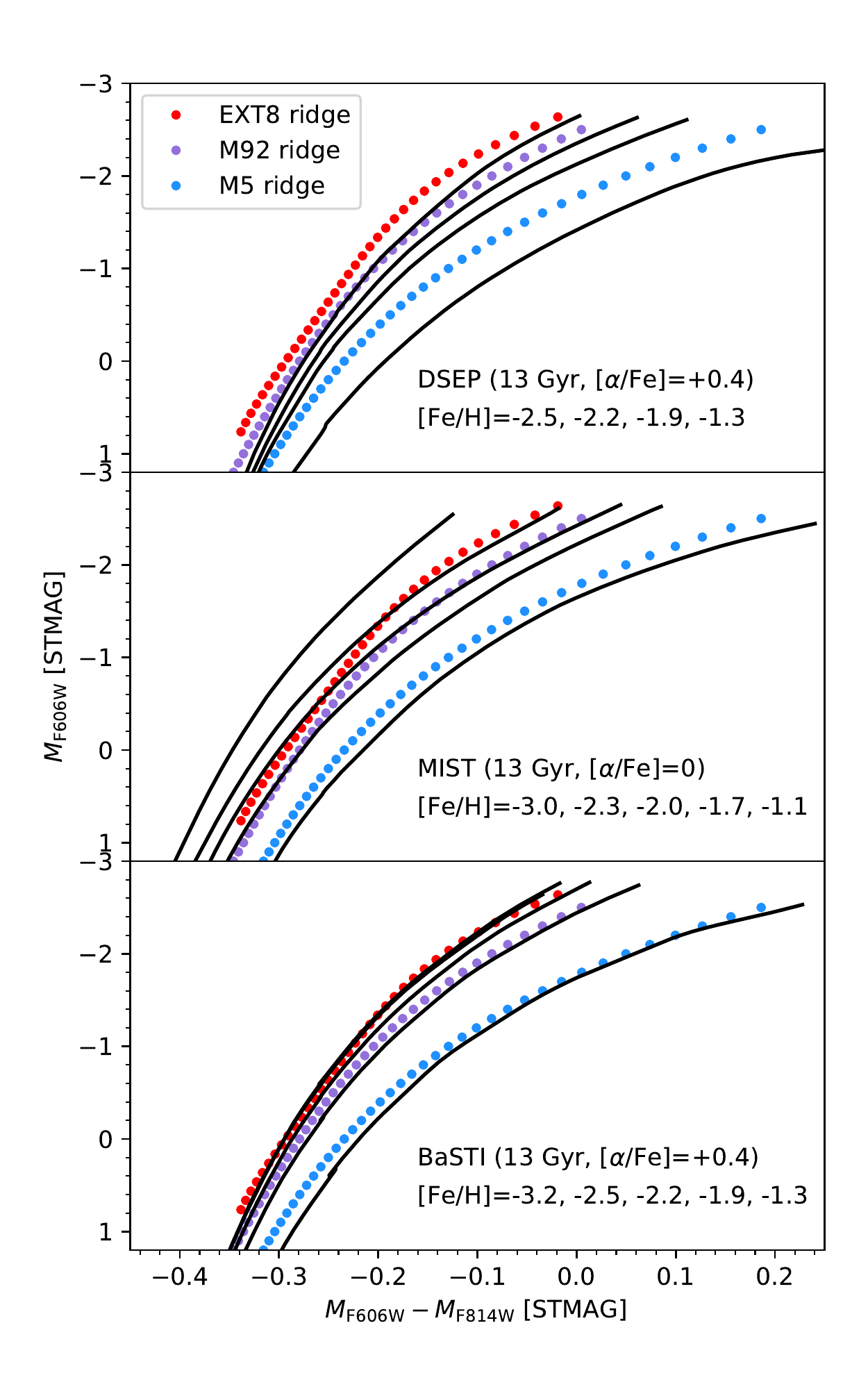}
\caption{\label{fig:isocmp}DSEP, MIST, and BaSTI isochrones for different metallicities compared with the RGB ridge lines for EXT8 (red dots), M92 (purple dots), and M5 (blue dots). The MIST isochrones are shown for higher $\mathrm{[Fe/H]}$ to compensate for their scaled-solar $\alpha$-element abundances.}
\end{figure}

\begin{figure}
\includegraphics[width=89mm]{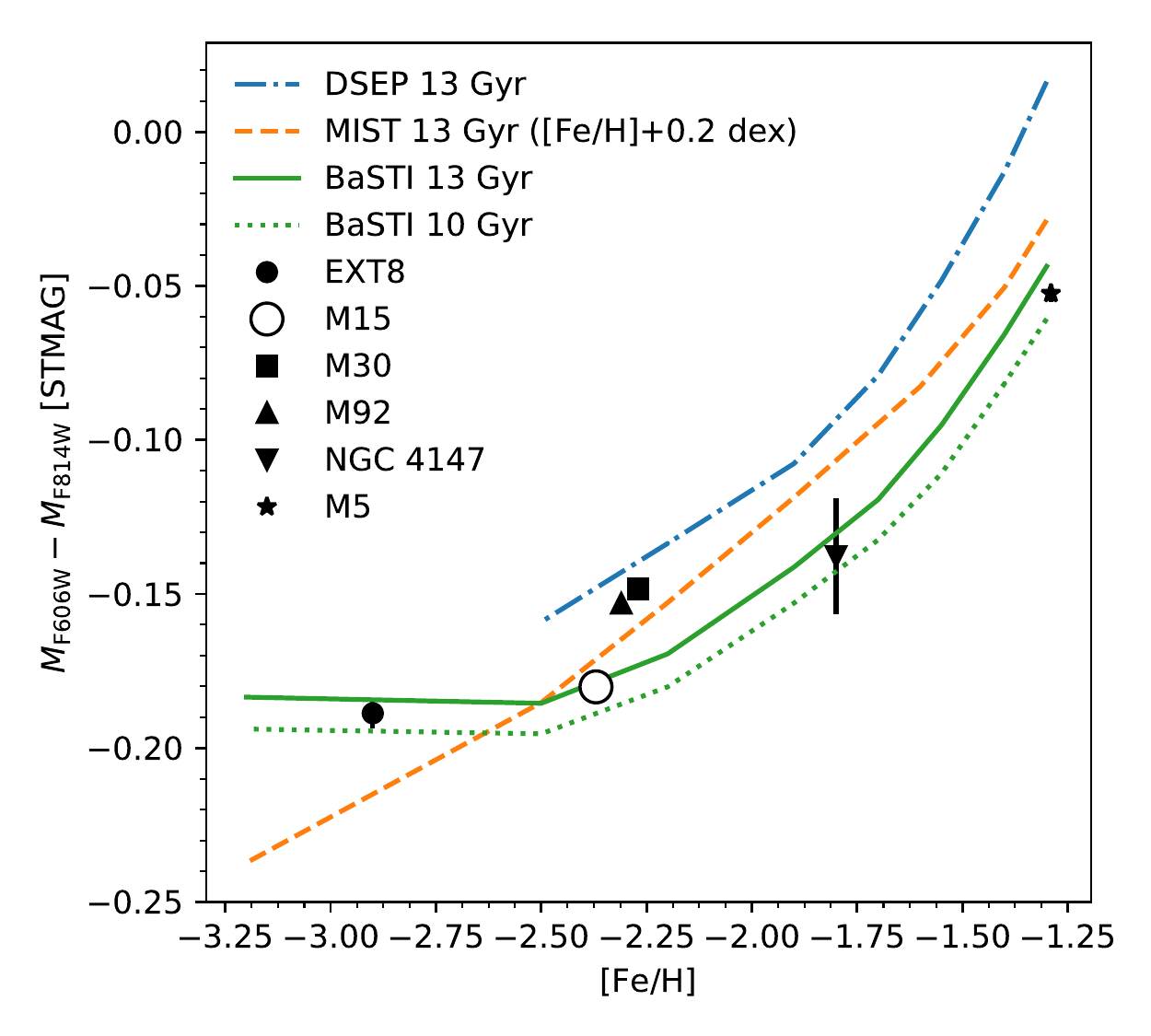}
\caption{\label{fig:fehvi}$M_\mathrm{F606W}-M_\mathrm{F814W}$ colour at $M_\mathrm{F606W}=-1.5$ versus $\mathrm{[Fe/H]}$  for DSEP, MIST, and BaSTI isochrones. The metallicities of the scaled-solar MIST isochrones are  offset by $-0.2$~dex for comparison with the $\alpha$-enhanced DSEP and BaSTI isochrones.
Also shown are the RGB colours for EXT8 and the Galactic GCs in Fig.~\ref{fig:cmdcmp}
(with M15 marked as an open symbol owing to doubts about the reddening correction).}
\end{figure}

We next compare the empirical results in the preceding section with predictions by theoretical isochrones.  Figure~\ref{fig:isocmp} shows the ridge lines for EXT8, M92, and M5 together with DSEP, MIST \citep{Choi2016}, and BaSTI isochrones.
For clarity the other ridge lines are omitted in this figure.
The BaSTI isochrones are shown for $\mathrm{[Fe/H]}=-3.2, -2.5, -2.2, -1.9$, and $-1.3$. The lowest iron abundance available for the DSEP isochrones is $\mathrm{[Fe/H]}=-2.5$ but they are otherwise shown for the same $\mathrm{[Fe/H]}$ values as the BaSTI isochrones. Unlike the DSEP and BaSTI isochrones, which are both shown for $\alpha$-enhanced composition ($[\alpha/\mathrm{Fe}]=+0.4$), the MIST isochrones are only available for scaled-solar composition. They are therefore shown for $\mathrm{[Fe/H]}=-3.0, -2.3, -2.0, -1.7$, and $-1.1$, which correspond to about the same total metallicities as those for which the BaSTI and DSEP isochrones are plotted \citep[see, for example,][]{Salaris1993}. All isochrones are shown for an age of 13 Gyr.

Apart from the fact that no set of isochrones perfectly matches all of the observed ridge lines, perhaps the most obvious difference is the much wider colour separation between the MIST isochrones at low metallicities compared to the BaSTI set. The $\mathrm{[Fe/H]}=-3.0$ and $-2.3$ MIST isochrones are separated by about 0.08~mag near the tip of the RGB, while the two corresponding BaSTI isochrones ($\mathrm{[Fe/H]}=-3.2$ and $-2.5$) are indistinguishable in the figure.
Overall, the BaSTI isochrones provide the best fits to the ridge lines for the range of magnitudes and metallicities shown here. The DSEP isochrones tend to be somewhat too red (as noted by \citealt{Vandenberg2013}), and the slopes of the most metal-poor MIST isochrones are shallower than the observed ridge lines. The difference in RGB slope between metal-poor MIST and BaSTI isochrones was previously noted by \citet{Hidalgo2018}.

A different way of comparing the various isochrones is shown in Fig.~\ref{fig:fehvi}, where the colours at a magnitude of $M_\mathrm{F606W}=-1.5$ are plotted as a function of metallicity for the different models. We also plot colours for 10-Gyr old BaSTI isochrones, which are about 0.01~mag bluer at a given metallicity compared to the 13-Gyr models. The effect of age differences of a few Gyr is thus relatively small, albeit not entirely negligible compared to the small RGB colour differences discussed here.
We include the colours of the RGB ridge lines for all GCs from Fig.~\ref{fig:cmdcmp} and we have drawn M15 with an open symbol as a reminder that its reddening correction may be more uncertain than for the other clusters. The larger error bar for NGC~4147 is due to its more sparsely populated RGB. EXT8 again has the bluest RGB ($M_\mathrm{F606W}-M_\mathrm{F814W}=-0.189\pm0.005$), followed closely by M15 
while M92 and M30 are 0.03--0.04~mag redder.
As in Fig.~\ref{fig:isocmp}, the different sets of isochrones reproduce different aspects of the observed trends, and all have various shortcomings. The BaSTI isochrones predict essentially no colour differences below $\mathrm{[Fe/H]}=-2.5$, which is in tension with the observed difference between EXT8 and M30/M92 (assuming that M15 is an outlier), but they otherwise reproduce the trend of colour vs.\ metallicity fairly well. Below $\mathrm{[Fe/H]}=-2.5$ the MIST models do show a continued trend towards bluer colours, but the predicted effect now appears too large compared with the observations. The DSEP isochrones are again redder than the MIST and BaSTI isochrones as well as the observed RGBs. 

An extensive discussion of the physical differences between the various isochrones would lead too far here. We note, for example, that the BaSTI isochrones are available for different assumptions about overshooting, mass loss, and diffusion, which can lead to differences of about 0.01~mag in the RGB colours at $\mathrm{[Fe/H]}=-2.9$. 
A more detailed exploration of how these and other differences in the underlying physics affect the behaviour of the models in the very metal-poor regime may be worthwhile.
In addition to differences in the physical parameters of the isochrones, significant differences in the observational plane can also arise depending on whether synthetic or empirical colour transformations are used \citep{Dotter2007}. 

When discussing RGB colour differences at the level of a few times 0.01~mag,
higher-order differences in the chemical abundance patterns may become relevant. In particular, we recall the peculiarly low magnesium abundance of EXT8. \citet{VandenBerg2012} investigated the effect of varying the abundances of specific elements and found that an increase of 0.4 dex in $\mathrm{[Mg/Fe]}$ made the upper RGB of an $\mathrm{[Fe/H]}=-2$ isochrone 
(at $M_V=-2$) about 0.05 mag redder in $V-I$ (read off of their Figure~16). 
We compared the $V-I$ colours of 13-Gyr, $\mathrm{[Fe/H]}=-1.9$ BaSTI isochrones with scaled-solar composition  and  $[\alpha/\mathrm{Fe}]=+0.4$  and found the difference to be about  0.04~mag, which is comparable  to  the  effect  of  varying Mg alone. While a change in Mg abundance can clearly have a non-negligible  effect  on  the  colours,  the  effect  is  likely  smaller  at the lower metallicity of EXT8. 
Given that the effect of changing the Mg abundance appears to be comparable to that of changing $[\alpha/\mathrm{Fe}]$ as a whole, an estimate may be made by comparing  the colours of scaled-solar and $\alpha$-enhanced BaSTI isochrones. 
For $\mathrm{[Fe/H]}=-2.9$ and an age of 13 Gyr we found a colour difference in $M_\mathrm{F606W}-M_\mathrm{F814W}$ of only 0.005 mag at $M_\mathrm{F606W}=-1.5$. 
Variations in helium abundance, as may be associated with multiple populations \citep{Milone2018}, will also affect the RGB colours. According to the BaSTI isochrones, an increase from $Y=0.247$ to $Y=0.30$ at $\mathrm{[Fe/H]}=-2.9$ makes the RGB about 0.004~mag bluer, so the effect is again expected to be small. While the different behaviour of the various isochrones at these low metallicities must be kept in mind, it appears unlikely that the detailed abundance patterns of EXT8 have a strong effect on the comparisons in the preceding paragraphs.

\subsubsection{Constraints on the metallicity spread}
\label{sec:zspread}

\begin{figure}
\includegraphics[width=89mm]{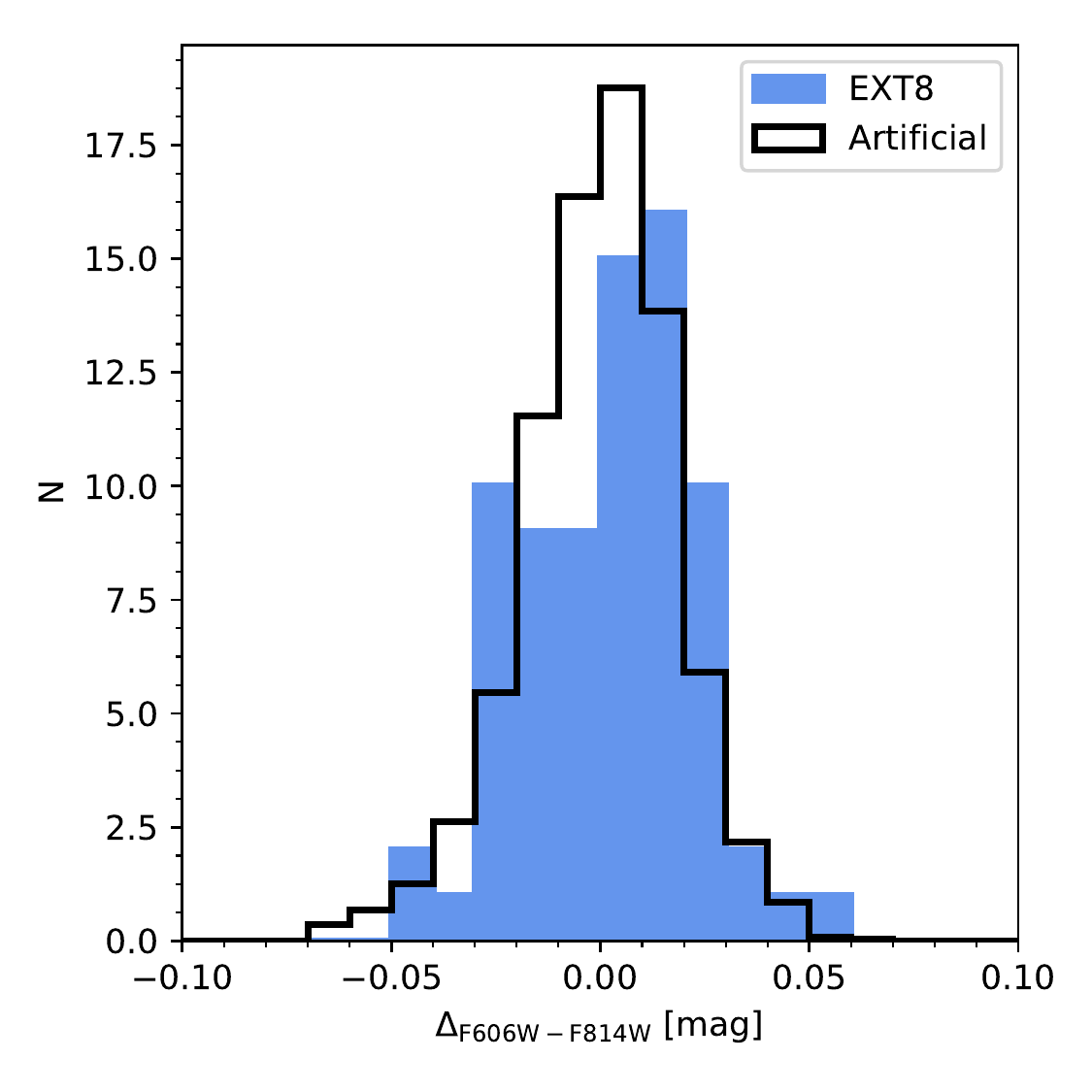}
\caption{\label{fig:dvihist}Distribution of RGB $\Delta_\mathrm{F606W-F814W}$ colour residuals with respect to ridge lines for EXT8 and the artificial star cluster. The artificial star distribution has been scaled to match the observed distribution. 
}
\end{figure}

To constrain any metallicity spread in EXT8, we compared the observed colour distribution of its RGB stars with the artificial star experiments. For this comparison we included stars down to a magnitude limit of $m_\mathrm{F606W} = 24.5$, located within the dashed lines in Fig.~\ref{fig:cmds}. At fainter magnitudes, the isochrones lie closer together and the observational errors increase, so that the sensitivity to a metallicity spread diminishes. 
Figure~\ref{fig:dvihist} shows the distributions of offsets $\Delta_\mathrm{F606W-F814W}$ between the measured colours of RGB stars and the ridge line for both the observed CMD and the artificial star experiments. For the observed CMD the dispersion is $\sigma_\mathrm{obs} = 0.0197$~mag (for 79 stars), and for the artificial stars we find $\sigma_\mathrm{synt} = 0.0185$~mag (1773 stars). The observed dispersion is thus slightly larger than indicated by the artificial star experiments.  Assuming that the $\Delta_\mathrm{F606W-F814W}$ offsets follow a normal distribution we can calculate the probability of finding a dispersion equal to or greater than $\sigma_\mathrm{obs}=0.0197$~mag when drawing 79 stars from a parent distribution with a dispersion of $\sigma_\mathrm{synt}=0.0185$~mag from the $\chi^2$ statistics. This probability is $P(\sigma_\mathrm{synt} > \sigma_\mathrm{obs}) = 23\%$, so there is no strong evidence of the difference being significant, even if we assume that the artificial star tests perfectly capture the observational uncertainties. As an additional test, we carried out a series of Monte-Carlo tests in which we resampled 79  $\Delta_\mathrm{F606W-F814W}$ offsets multiple times from the artificial star photometry and recomputed the dispersion.  The dispersions exceeded the observed value of 0.0197~mag in about 21\% of the realisations, in good agreement with the $\chi^2$-based estimate. 
The artificial star CMD, which was based on DSEP isochrones, does not include any potential contamination of the RGB by AGB stars. We repeated the artificial star experiments using BaSTI isochrones, which do include the AGB, and found that the dispersion of the $\Delta_\mathrm{F606W-F814W}$ distribution indeed increased slightly to $\sigma_\mathrm{synt} = 0.0219$~mag, now slightly broader than $\sigma_\mathrm{obs}$.
We are led to conclude that the observed colour spread on the RGB can be fully explained by the photometric uncertainties.

We next quantify how large a metallicity spread can be ruled out. This is complicated by the significantly different colour-metallicity relations predicted for the RGB by the various isochrones, and we therefore started with a more model-independent comparison based on colour spreads alone. 
To this end we randomly added normally distributed colour offsets to the $\Delta_\mathrm{F606W-F814W}$ offsets from the artificial star tests. We then compared the resulting colour distributions with the observations to find out whether the broadened artificial star colour distribution was still consistent with being at least as narrow as the observed colour distribution.
For an intrinsic colour spread of $\sigma_i = 0.005$~mag, the total dispersion of the artificial star colour distribution was $0.0193$~mag, still slightly narrower than the observed distribution, and the probability of finding a dispersion smaller than or equal to the observed dispersion was found to be $P(\sigma_\mathrm{synt} < \sigma_\mathrm{obs}) = 67\%$, implying that this is certainly allowed. For larger intrinsic colour dispersions, the corresponding probabilities were found to be 
$P(\sigma_\mathrm{synt} < \sigma_\mathrm{obs}) = 27\%$ (for $\sigma_i = 0.010$~mag), 
$P(\sigma_\mathrm{synt} < \sigma_\mathrm{obs}) = 2.4\%$ ($\sigma_i = 0.015$~mag), and
$P(\sigma_\mathrm{synt} < \sigma_\mathrm{obs}) = 0.04\%$ ($\sigma_i = 0.020$~mag). 
 An intrinsic colour dispersion greater than 0.015~mag is therefore ruled out at about the two sigma level. When basing the analysis on the BaSTI artificial star experiments, this limit is reduced to about 0.010~mag. 

To map these colour dispersions to metallicity dispersions, we used the isochrones for guidance. 
Interpolating between the BaSTI $\mathrm{[Fe/H]}=-2.5$ and $\mathrm{[Fe/H]}=-3.2$ isochrones, we found an RGB colour of $M_\mathrm{F606W}-M_\mathrm{F814W} = -0.184$ for $\mathrm{[Fe/H]}=-2.9$ at a reference magnitude of $M_\mathrm{F606W}=-1.5$ (Fig.~\ref{fig:fehvi}). The $\mathrm{[Fe/H]}=-2.2$ isochrone has $M_\mathrm{F606W}-M_\mathrm{F814W} = -0.170$, about $0.014$ mag bluer. This is almost the same as the 2$\sigma$ limit on the colour dispersion, so that a metallicity dispersion larger than the difference between the two isochrones, or 0.7~dex, is then ruled out at about the 2$\sigma$ confidence level. However, this assumes that a symmetric metallicity distribution maps to a symmetric colour distribution, which is probably not the case according to the BaSTI isochrones - indeed, the lower bound on $\mathrm{[Fe/H]}$ is essentially unconstrained since there are no BaSTI isochrones with $\mathrm{[Fe/H]}<-3.2$, but we recall that the RGB colours remain nearly constant between $\mathrm{[Fe/H]}=-2.5$ and $\mathrm{[Fe/H]}=-3.2$.
If we instead use the MIST isochrones, interpolation gives $M_\mathrm{F606W}-M_\mathrm{F814W} = -0.219$ for $\mathrm{[Fe/H]}=-2.7$ (accounting for the scaled-solar composition). Owing to the wider colour separation between these isochrones, a metallicity dispersion of about 0.2~dex maps to a colour dispersion of about 0.013~mag in both directions, again not far from the 0.015~mag upper limit.  Were we to rely on the MIST isochrones, the two-sigma upper limit on the metallicity dispersion would therefore be about 0.2~dex. 
As an alternative to these model comparisons, we may take the $M_\mathrm{F606W}-M_\mathrm{F814W}$ colour difference of $\sim0.03$~mag between the RGB of EXT8 and the metal-poor Galactic GCs (Sect.~\ref{sec:rgbemp}) 
as indicative of a metallicity difference of $\sim0.6$~dex. The allowed $\sim0.015$~mag colour dispersion would then translate to a metallicity dispersion of  $\sigma_\mathrm{[Fe/H]}\approx0.3$~dex.
In summary, the main uncertainty in quantifying the limit on the metallicity spread is the colour--metallicity relation for the metal-poor RGB stars.

\subsubsection{The tip of the RGB and the distance to EXT8}
\label{sec:trgb}

The $I$-band luminosity of the tip of the RGB (TRGB) is only weakly sensitive to metallicity and is commonly used as a distance indicator \citep{Lee1993}. Although the TRGB is not well sampled in our CMD of EXT8, we can use the brightest RGB star (marked with a circle in Fig.~\ref{fig:img}) to obtain a consistency check of the adopted distance, assuming this star is indeed a member of EXT8. The star is located well outside the crowded central region of EXT8 and has reddening-corrected magnitudes of $m_\mathrm{F606W,0} = 21.71\pm0.01$ and $m_\mathrm{F814W,0} = 21.66\pm0.01$. Transforming these magnitudes to the Johnson--Cousins system \citep{Harris2018} gives $m_{I,0} = 20.40$ and $(V-I)_0 = 1.32$, or $M_I = -4.07$ for the assumed distance. Recent calibrations of the TRGB magnitude range between $M_{I,\mathrm{TRGB}} = -3.99$ and $M_\mathrm{I,\mathrm{TRGB}} = -4.16$ \citep{Jang2017,Capozzi2020}. While these calibrations are given for somewhat redder TRGB colours, the corresponding $M_{I,\mathrm{TRGB}}$ magnitudes would be less than 0.01~mag fainter at the colour of the EXT8 TRGB \citep{Jang2017}.
The photometry of the brightest RGB star is thus consistent with EXT8 being located at the assumed distance. 

However, given that the upper part of the RGB is not well sampled, it is quite possible that we have not detected the true tip of the RGB, in which case the agreement would be less good and EXT8 might be somewhat closer than assumed. Indeed, our CMD only contains 14 stars within 1 magnitude of the brightest RGB star, while a reliable estimate of the TRGB magnitude requires at least $\sim50$ stars in this range \citep{Madore1995}. 
The TRGB then appears consistent with a slightly smaller distance, as also suggested from isochrone fitting to the HB, but does not favour the larger distance modulus suggested by the comparison of the HB brightness with the empirical relation of \citet{Dotter2010} (Sect.~\ref{sec:hb}). 

If EXT8 is indeed closer than assumed, this would make the RGB redder at a given absolute magnitude. If we decrease the distance modulus by 0.1~mag and repeat the comparison of colour offsets from Sect.~\ref{sec:rgbemp}, we find a shift of 0.01~mag towards the red at fixed $M_\mathrm{F606W}$.
The RGB of EXT8 thus remains bluer than all of the Galactic GCs except M15 for any reasonable range of distances allowed by the data. It would take a shift of about 0.3~mag in the distance modulus (100 kpc in the distance) to make the EXT8 RGB as red as those of M92 and NGC~5466, but such a large shift would make the HB clearly too faint.

\subsection{The horizontal branch}
\label{sec:hb}

\begin{figure}
\includegraphics[width=89mm]{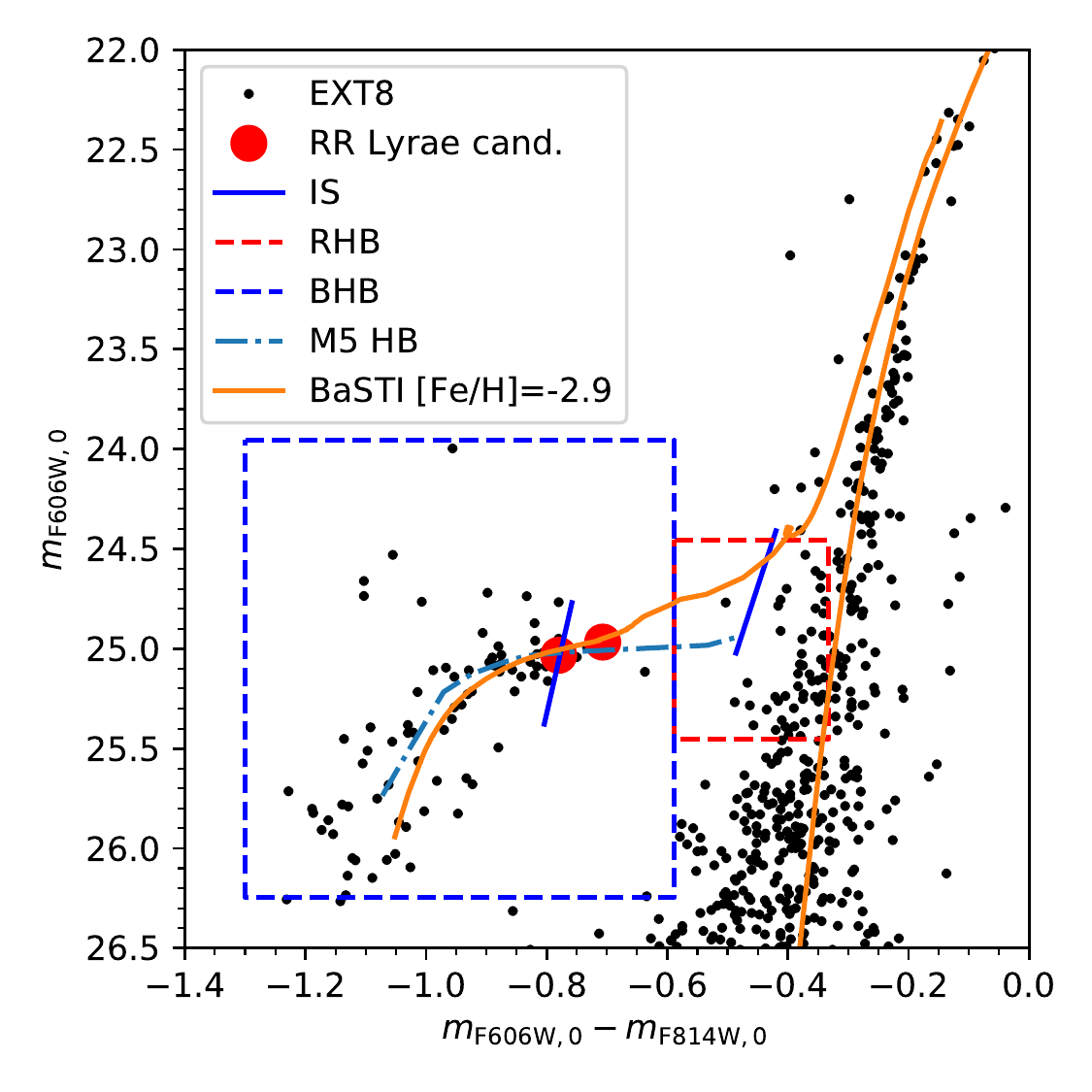}
\caption{\label{fig:ext8fit}Colour--magnitude diagram of EXT8 overlaid with the HB of M5 and an $\mathrm{[Fe/H]}=-2.9$ BaSTI isochrone with an age of 13~Gyr
(in both cases shifted vertically to match up with EXT8). 
The boxes show the selection regions for BHB and RHB stars,
and the slanted lines show the IS boundaries.
The two RR Lyrae candidates are marked with larger red circles.}
\end{figure}

\subsubsection{General morphology and HB magnitude}
\label{sec:hbmorph}

Classical parameters to quantify HB morphology include the \citet{Mironov1972} index, $\mathrm{MI} \equiv B/(B+R)$, and $\mathrm{HBR} \equiv (B-R)/(B+V+R)$ \citep{Lee1994}, where $B$, $V$, and $R$ are the numbers of stars bluer than the IS, in the IS, and redder than the IS, respectively. The apparent simplicity of these definitions is somewhat deceptive and complications arise because 1) not all HB stars may be detected, 
in particular if a blue tail is present and extends below the detection limit, 2) RR Lyrae stars may scatter out of the IS if the pulsation periods are incompletely sampled, and 3) red HB stars may be confused with RGB stars. 
For GCs at the distance of M31, where RR Lyrae identification is challenging,
a simplified version of the Mironov index (SMI) can be defined by splitting the HB in two, roughly in the middle of the IS \citep{Rich2005}. Here we adopt the definition of the SMI from \citet{Perina2012}, counting red HB (RHB) stars as stars having $0.5 < (V-I)_0 < 0.8$ and lying within $\pm0.5$~mag of the $V$ magnitude of the HB ($V_\mathrm{HB}$) and blue HB (BHB) stars as stars having $(V-I)_0 < 0.5$ and with $V_\mathrm{HB}-1 < V_0 < 26$. Converted to WFC3 STMAG magnitudes, this corresponds to $-0.59 < M_\mathrm{F606W}-M_\mathrm{F814W} < -0.33$ for the colour limits of the RHB and $m_\mathrm{F606W,HB,0}-1 < m_\mathrm{F606W,0} < 26.25$ for the magnitude limits of the BHB.

To determine the magnitude of the HB, $m_\mathrm{F606W,HB,0}$, we followed a procedure similar to that described in \citet{Dotter2010}. Noting that the HB of M5 spans a wide range of colours, they shifted the HBs of other GCs relative to the M5 HB and then determined $m_\mathrm{F606W,HB}$ as the $m_\mathrm{F606W}$ magnitude of the horizontal part of the shifted M5 HB. For M5 itself they found $m_\mathrm{F606W,HB,VEGA}=14.90\pm0.05$ in the VEGAMAG system. We defined a fiducial HB sequence for M5 using the same ACSGCS data as \citet{Dotter2010} (Fig.~\ref{fig:cmdcmp}).
We calculated the median $m_\mathrm{F606W,0}$  magnitude in 0.05~mag bins of $M_\mathrm{F606W}-M_\mathrm{F814W}$ colour for  $-1.10 < M_\mathrm{F606W}-M_\mathrm{F814W} < -0.45$, skipping the range $-0.8 < M_\mathrm{F606W}-M_\mathrm{F814W} < -0.5$ where RR Lyrae variables cause a large scatter in colour and magnitude in the M5 CMD. For the resulting fiducial sequence we found $m_\mathrm{F606W,HB,0} = 15.07$ at $m_\mathrm{F606W,0}-m_\mathrm{F814W,0}=-0.5$. Converting to VEGAMAG magnitudes and adding back the foreground extinction, this becomes $m_\mathrm{F606W,HB,VEGA}=14.92$, matching the value found by \citet{Dotter2010} within 0.02~mag. 

Figure~\ref{fig:ext8fit} shows the reddening corrected CMD of EXT8 with the shifted M5 HB sequence and the BHB and RHB boxes overplotted. A shift of $\Delta M_\mathrm{F606W} = 9.88\pm0.03$~mag was required to match the fiducial M5 HB to the EXT8 HB, calculated for colours in the range $-1.0 < M_\mathrm{F606}-M_\mathrm{F814W} < -0.8$. This gave $m_\mathrm{F606W,HB,0} = 24.96\pm0.03$~mag. 
While, as noted above, a few RHB stars may be present in the CMD, it is obvious that the vast majority of the 55 stars located within the RHB box are RGB stars. In a few similar cases, \citet{Perina2012} were able to detect a peak in the RGB luminosity function (LF) near the expected level of the RHB and they could then statistically subtract the underlying RGB population via interpolation in the LF. However, no such excess of RHB candidates was detected for EXT8. The BHB box contains 82 stars, while the field region (Fig.~\ref{fig:cmds}) contains 18 stars in the BHB box. Most of the remaining 64 stars are likely BHB stars.
If EXT8 contains no RHB stars at all, we simply get $\mathrm{MI} = \mathrm{SMI} = 1$, and allowing for 5 RHB stars would give $\mathrm{SMI}=0.93$. A crude estimate would then be $0.9 \la \mathrm{SMI} \leq 1$.

The magnitude of the HB is well known to depend on metallicity, in the sense that more metal-poor GCs have brighter HBs \citep{Lee1990}. \citet{Dotter2010} found $M_\mathrm{F606W,HB,VEGA} = (0.227\pm0.011)\mathrm{[Fe/H]} +0.802\pm0.020$ for Milky Way GCs, and similar relations have been found for M31 GCs \citep{Rich2005,Federici2012}.
Extrapolated to $\mathrm{[Fe/H]}=-2.9$, the relation in Dotter et al.\ gives $M_\mathrm{F606W,HB,VEGA} = +0.14\pm0.04$ or $M_\mathrm{F606W, HB} = +0.39\pm0.04$ (STMAG). 
The value measured for EXT8 from the shifted M5 HB is $M_\mathrm{F606W, HB} = +0.49\pm0.03$ at the assumed distance, 
about 0.1~mag fainter than the expected value. However, at $\mathrm{[Fe/H]}=-2.4$ the predicted magnitude is $M_\mathrm{F606W, HB} = +0.50$, which is nearly identical to the value measured for EXT8. If EXT8 is indeed located at the assumed distance, this might suggest that the $M_\mathrm{F606W, HB}$ vs.\ $\mathrm{[Fe/H]}$ relation flattens below $\mathrm{[Fe/H]}=-2.4$. Alternatively, taking the predicted $M_\mathrm{F606W, HB}$ for $\mathrm{[Fe/H]}=-2.9$ at face value, the implied distance modulus would be $(m-M)_0 = 24.57\pm0.05$, taking into account the uncertainties on both the predicted and measured HB magnitudes. 

In the same way that the M5 HB fiducial sequence was shifted to match the EXT8 HB, we determined the vertical shift that gave the best fit to a theoretical isochrone. Figure~\ref{fig:ext8fit} includes a 13 Gyr BaSTI isochrone interpolated to $\mathrm{[Fe/H]}=-2.9$, for which we found a shift of $(m-M)_0 = 24.43\pm0.03$~mag. This is consistent with the distance modulus $(m-M)_0 = 24.47$ that corresponds to the assumed distance of M31, but we note that this small shift is in the opposite sense to that obtained from the \citet{Dotter2010} relation and would thus lead to a larger discrepancy between the measured HB magnitude and the prediction. The independent constraint on the distance obtained from the tip of the RGB also appears to support a distance to EXT8 no greater than the assumed mean distance of M31 (Sect.~\ref{sec:trgb}). The shape of the HB predicted by the BaSTI models is somewhat different from the M5 fiducial sequence, probably due to the higher metallicity of M5.  \citet{Pietrinferni2021} show that the HBs of the BaSTI isochrones match GCs of different metallicities well, and also reproduce the trend of HB luminosity versus metallicity. Part of the discrepancy between the predicted and observed $M_\mathrm{F606W, HB}$ magnitudes might then be the result of applying the M5 HB at the low metallicity of EXT8. 

Our estimates of $(m-M)_0$ thus lie in the range between $24.43\pm0.03$ (about 770 kpc) and $24.57\pm0.05$ (824 kpc), with a preference for distances towards the lower end of the range. 
This is then consistent with a distance modulus of about 24.46 for M31 \citep{Stanek1998,Conn2012} but still allows a considerable range of line-of-sight separations between the centre of M31 and EXT8. 
It is also worth noting that the distance of M31 itself is subject to systematic uncertainties, depending on the choice of distance indicators, where a shorter distance modulus of $(m-M)_0 = 24.38\pm0.06$ (752 kpc) has been deduced from Cepheids \citep{Riess2012}. The range of 3-D separations between EXT8 and the centre of M31 could then lie between the projected value of 27~kpc and up to as much as 70~kpc, although the latter would require a perhaps somewhat unlikely conspiracy of uncertainties.

\subsubsection{The HB in the ultraviolet}
\label{sec:hbuv}

\begin{figure}
\includegraphics[width=89mm]{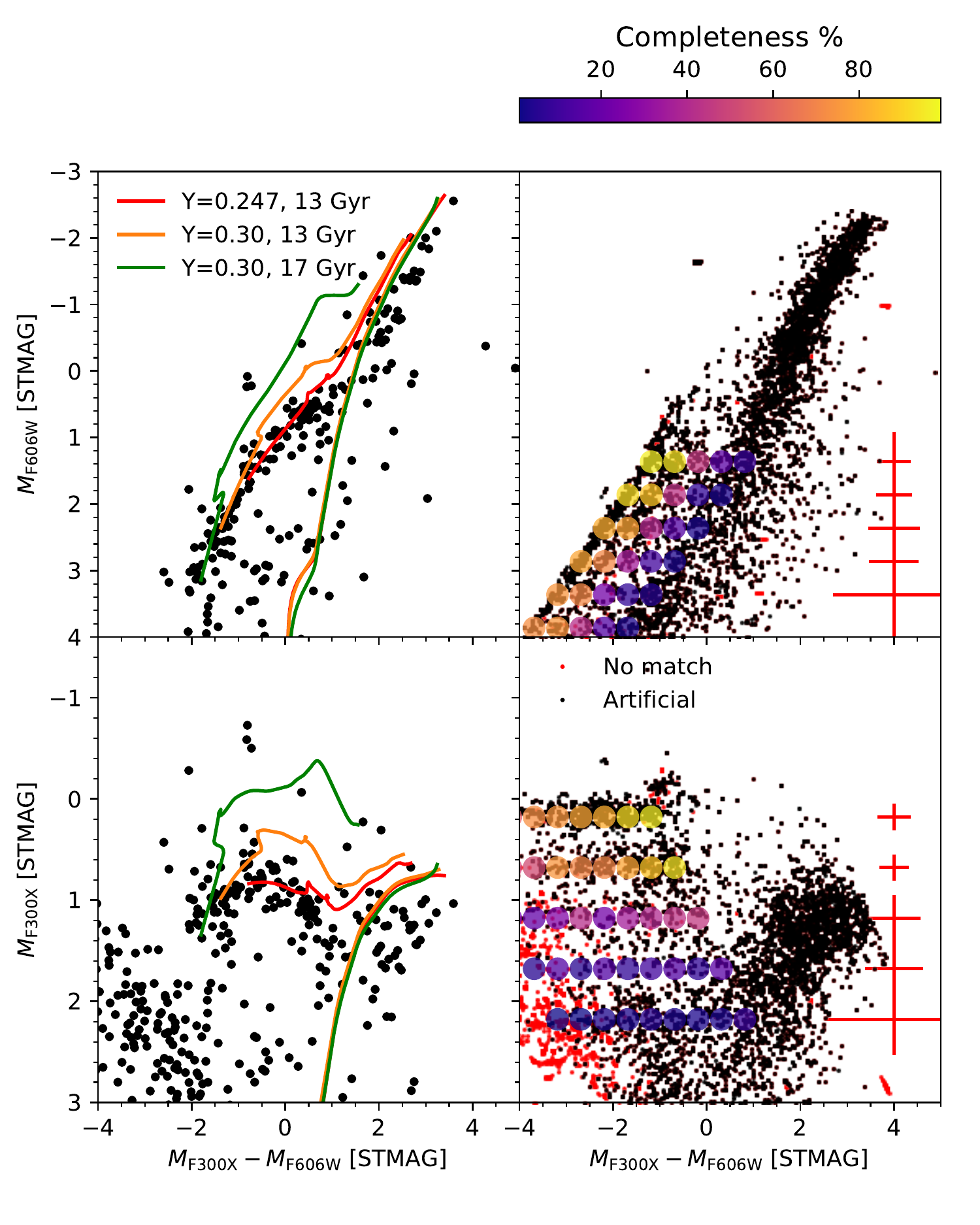}
\caption{\label{fig:cmduvi}UV photometry for EXT8 (left-hand panels) and artificial star experiments (right-hand panels). Isochrones are shown for $\mathrm{[Fe/H]}=-3.2$ and helium abundances and ages as indicated in the legend.
The red symbols are sources that do not have a match in the input artificial star catalogue, most of which are spurious. The detection completeness is indicated by the coloured circles. Error bars are shown for stars with typical HB colours ($-2 < M_\mathrm{F300X}-M_\mathrm{F606W} < +1$).
}
\end{figure}

Based on the artificial cluster CMD and completeness tests (Fig.~\ref{fig:cmds}), HB stars in EXT8 are expected to be detectable well below the boundaries of the BHB box. 
However, at magnitudes fainter than $M_\mathrm{F606W}\approx+2$, stars on the sub-RGB and near the MSTO (including any blue stragglers, if present in significant numbers in EXT8) with large photometric errors start scattering into the region of the optical CMD occupied by the HB, making the identification of potential extended BHB stars more ambiguous.
While some stars with colours consistent with an extended BHB  are visible down to $M_\mathrm{F606W}\approx+3$ in the optical CMD, it is not clear from Fig.~\ref{fig:cmds} or Fig.~\ref{fig:cmdcmp} how many of them, if any, are actually HB stars. We therefore now turn to the analysis of the F300X images, which can help separate the populations in this part of the CMD. 

In Fig.~\ref{fig:cmduvi} (left column) we show the CMDs for stars detected in the F300X image while the right column shows the results of the artificial star tests. In addition to the HB and the brighter RGB stars, which have comparable F300X magnitudes, a small number of stars with magnitudes $M_\mathrm{F300X} < 0$ are prominently visible above the HB. These stars are also visible in the optical CMDs, where they lie about a magnitude above the blue part of the HB (Fig.~\ref{fig:ext8fit}). These are possibly ``UV-bright" post-HB or post-AGB stars \citep{Zinn1972,Harris1983,Moehler2019,Bond2021a}. Although some field contamination is possible in this region of the CMD (Fig.~\ref{fig:cmds}), the UV-bright candidates appear spatially associated with EXT8 with four of them located at radii $3\arcsec < R < 7\arcsec$ and the fifth at about 15\arcsec. These stars are marked on the colour image in Fig.~\ref{fig:img}. 

As in Fig.~\ref{fig:cmds}, the coloured circles in the right-hand panels of Fig.~\ref{fig:cmduvi} show the completeness levels at various locations in the CMD according to the colour scale at the top. The black points in the right-hand panels are sources from the input artificial star catalogue that were recovered in the F300X photometry, while red points indicate sources that have no counterpart in the input catalogue and are mostly spurious detections. 
The errors depend on both the colours and magnitudes of the stars, but we have included error bars for stars with typical HB colours. As in Fig.~\ref{fig:cmds}, the error bars are based on the artificial star experiments.
It is clear that the detection here strongly favours stars on the HB and we can now see that the HB in EXT8 does in fact extend at least to $M_\mathrm{F606W}\approx+3$. 
The completeness in F300X is about 50\% at the level of the detected HB stars, but for colours bluer than $M_\mathrm{F300X}-M_\mathrm{F606W} \approx-2$ the HB starts becoming fainter also in F300X and the detection efficiency rapidly drops. Deeper observations would be required to ascertain whether the HB extends even further and includes a more extreme blue tail as seen in some Galactic GCs \citep{FusiPecci1993}.
A number of even fainter sources are visible in the lower left-hand corner of the $M_\mathrm{F300X}$ vs. $M_\mathrm{F300X}-M_\mathrm{F606W}$ CMD, but comparison with the artificial cluster CMD shows that these are mostly spurious detections. This was further confirmed by an inspection of the spatial distribution of these sources, which showed them to be mostly random noise peaks, uniformly distributed across the detector.
Adopting a more conservative detection threshold for the \texttt{ALLFRAME} photometry would lead to fewer such spurious detections, but would also decrease the detection efficiency for real sources. 

As an alternative to the SMI index considered in Sect.~\ref{sec:hbmorph} we can now consider the $\Delta(V-I)$ index, defined as the difference between the median $(V-I)$ colours of the HB stars and RGB stars at the level of the HB \citep{Dotter2010}. The median colour of the 50 RGB stars in the range $m_\mathrm{F606W,HB} -0.25 < m_\mathrm{F606W} < m_\mathrm{F606W,HB}  + 0.25$ is $(M_\mathrm{F606W} - M_\mathrm{F814W})_\mathrm{RGB} = -0.33$. In addition to the 82 stars in the BHB box (Fig.~\ref{fig:ext8fit}), we counted 48 stars with $m_\mathrm{F606W,0} > 26.25$ and $M_\mathrm{F300X} - M_\mathrm{F606W} < -1$ in Fig.~\ref{fig:cmduvi}. The median colour of the total sample of 130 HB stars is $M_\mathrm{F606W} - M_\mathrm{F814W} = -1.10$, assuming that all HB stars with $m_\mathrm{F606W,0} > 26.25$ have bluer colours than those in the BHB box.
However, this still does not take into account the $\sim18$ field contaminants in the BHB box, which are likely found mostly near the red boundary of the HB (Sect.~\ref{sec:cmd0}). If we exclude the 18 reddest HB stars from the sample, the median HB colour shifts to $M_\mathrm{F606W} - M_\mathrm{F814W} = -1.14$. The median colour difference is then $\Delta(\mathrm{F606W}-\mathrm{F814W}) = 0.81$~mag or $\Delta(V-I) = 0.95$~mag. This is probably still an underestimate of $\Delta(V-I)$ since we are likely missing a significant number of stars in the tail of the HB. Our estimate is therefore best expressed as a lower limit, $\Delta(V-I) > 0.95$, and the HB morphology as EXT8 might be as blue as the bluest HBs measured by \citet{Dotter2010}.

To illustrate the effect of variations in helium abundance on the HB, Fig.~\ref{fig:cmduvi} includes $\mathrm{[Fe/H]}=-3.2$ BaSTI isochrones with $Y=0.247$ and $Y=0.30$ for an age of 13~Gyr. It is clear that the EXT8 HB extends beyond the bluest extent ($T_\mathrm{eff}\approx11400$~K) of the $Y=0.247$ isochrone. 
Stars with an increased helium content evolve faster and are therefore expected to arrive at the tip of the RGB (and then on the zero-age horizontal branch, ZAHB) with lower masses, and hence with less massive envelopes and correspondingly higher temperatures.
For the BaSTI models in Fig.~\ref{fig:cmduvi} the masses at the tip of the RGB are $M = 0.705 \,  M_\odot$ for $Y=0.247$ and $M = 0.638 \, M_\odot$ for $Y=0.30$, respectively. The HB models indeed extend further to the blue for $Y=0.30$, reaching $T_\mathrm{eff} \approx 14500$~K, but still do not quite reach the most extreme HB stars that we detect in EXT8. 
A further increase in helium abundance would lead to still lower ZAHB masses and a correspondingly bluer HB, and the HB morphology might thus suggest some tentative evidence for the presence of relatively helium-rich stars in EXT8, perhaps related to the very low magnesium abundance inferred from the integrated-light spectrum. 

It should be stressed, however, that many other ``second parameters'' besides helium abundance and age have been proposed \citep{FusiPecci1997,Catelan2009} and HB modelling in general is highly sensitive to the amount of mass loss along the RGB, which remains uncertain \citep{Salaris2016}. The effect of an increased mass loss can be mimicked by selecting an older isochrone: for an age of 17~Gyr, $\mathrm{[Fe/H]}=-3.2$ and $Y=0.30$, the ZAHB mass is $M = 0.577 \, M_\odot$, which produces an HB that extends about as far as the faintest HB stars detected in EXT8 and reaches temperatures of $T_\mathrm{eff}\approx18000$~K.

\subsubsection{Variability}
\label{sec:rrlyr}

\begin{figure*}
\centerline{\includegraphics[width=150mm]{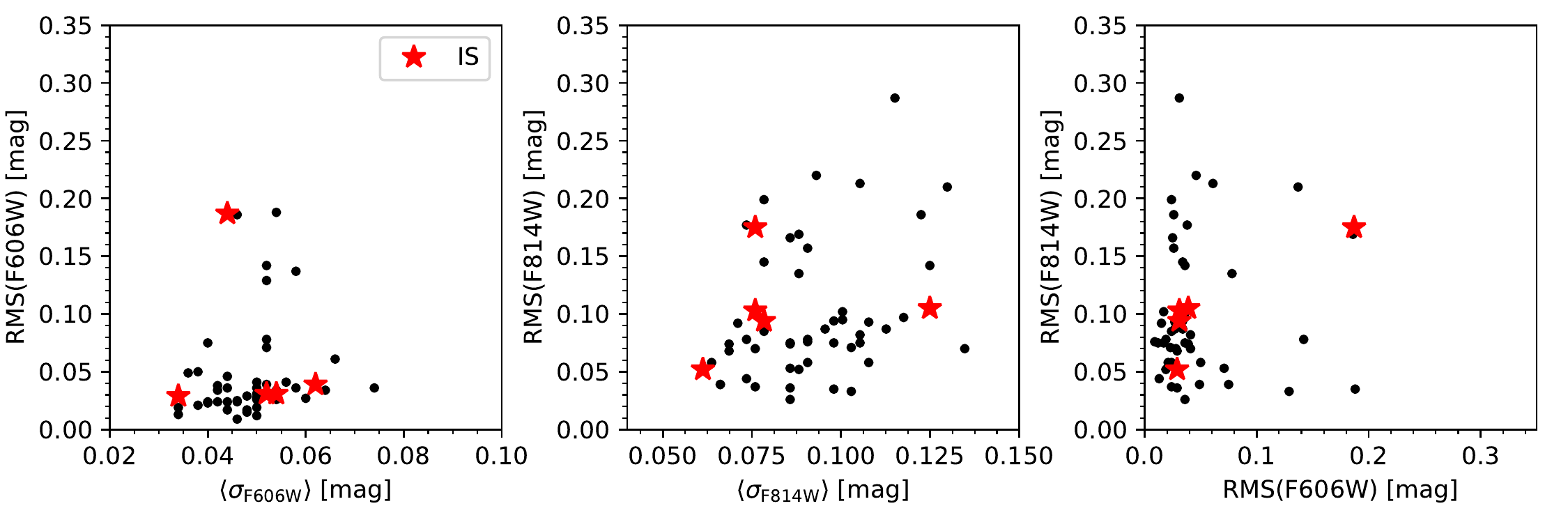}}
\caption{\label{fig:err_sig}Frame-to-frame rms dispersion of the magnitude measurements versus mean photometric errors for HB stars in EXT8. The asterisks indicate the five stars located within the IS (Fig.~\ref{fig:cmdcmp}).}
\end{figure*}

\begin{figure}
\includegraphics[width=89mm]{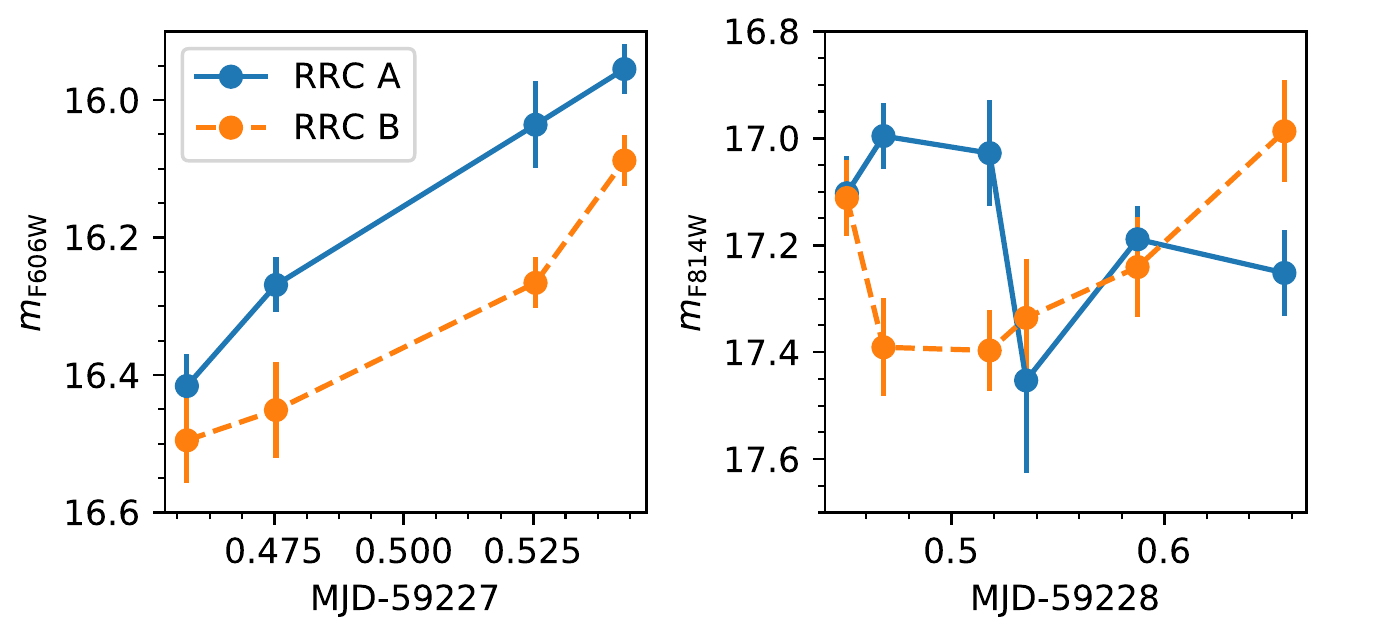}
\caption{\label{fig:rrc}Light curves for the two RR Lyrae candidates. }
\end{figure}

The F606W and F814W observations were obtained over 2 hours and 5 hours, respectively, and therefore do not fully sample typical RR Lyrae pulsation periods of 0.3--0.8 days \citep{Oosterhoff1939,Carney1992,Hoffman2021}. Nevertheless, some variation on time scales of a few hours might be detectable, and as noted above the CMD does show a few stars located within the boundaries of the IS. Most of these are located near the blue edge and could be non-variables scattering into the IS from the part of the HB located just outside the IS. Conversely, some genuine variables might scatter out of the IS.

To look for evidence of variability among the HB stars, we plotted the rms variation of the F814W and F606W magnitudes measured on the individual \texttt{\_flc} images versus the mean errors reported by \texttt{ALLFRAME} for stars with $-1.5 < M_\mathrm{F606W}-M_\mathrm{F814W} < -0.50$ and with $-0.5 < M_\mathrm{F606W} < +1.0$ (Fig.~\ref{fig:err_sig}). 
We recall that the \texttt{ALLFRAME} errors on average reproduce the random errors on the measurements well (Sect.~\ref{sec:artex}), so that a significantly larger rms variation than expected based on the errors might indicate a genuine change in brightness from one frame to another.
The five stars located within the IS are marked with asterisks. There is no strong evidence that the distribution of these stars is different from that of the HB stars in general in these plots, but one of them stands out by having a relatively large rms variation in both F606W and F814W. 
Another star, located just outside the blue edge of the IS, has a similarly large variation, $\mathrm{rms}>0.15$~mag in both filters. These two stars may be considered RR Lyrae candidates and are marked with dashed circles in Fig.~\ref{fig:img} and with large red circles in Fig.~\ref{fig:ext8fit}.
The outer star, which we refer to as RR Lyrae candidate A (RRC A), is located at a projected separation of about $19\arcsec$ from the centre of EXT8. This is the star located furthest to the red in the CMD, inside the IS. RRC~A is unaffected by crowding and we confirmed the variations in magnitude measured by \texttt{ALLFRAME} by remeasuring the star with the \texttt{imexamine} task in \texttt{IRAF}. The inner star, RRC B, is located about $3\farcs2$ from the centre of EXT8. The light curves for the two candidates are shown in Fig.~\ref{fig:rrc}. Both stars show a relatively smooth variation in both filters, consistent with variability on time scales expected for RR Lyrae variables, although the coverage is too incomplete to determine periods. 
It is again worth recalling that even if these stars are genuine RR Lyrae variables, some field contamination is expected in this region of the CMD. 

The right-hand panel in Fig.~\ref{fig:err_sig} shows several other stars with a relatively large rms variation in F814W, $\mathrm{rms(F814W)} \ga 0.15$~mag, but smaller variations in F606W. We found the distribution of these stars in the CMD to be similar to the general distribution of HB stars. Most of them tend to have relatively large errors (middle panel) and are located within a radius of 5\arcsec\ from the centre of EXT8 where crowding increases the photometric uncertainties.

\subsection{A UV-bright source near the centre of EXT8}
\label{sec:uvbright}

\begin{figure}
\begin{minipage}{44mm}
F300X \\
\includegraphics[width=44mm]{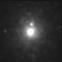}
\end{minipage}
\begin{minipage}{44cm}
Residuals \\
\includegraphics[width=44mm]{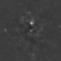}
\end{minipage}
\begin{minipage}{44mm}
F606W \\
\includegraphics[width=44mm]{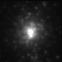}
\end{minipage}
\begin{minipage}{44mm}
Residuals\\
\includegraphics[width=44mm]{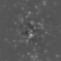}
\end{minipage}
\caption{\label{fig:uvc}The central $2.4^{\prime\prime}\times2.4^{\prime\prime}$ of EXT8. The left-hand panels show the F300X and F606W images and the right-hand panels show the residuals after subtraction of the best-fitting model profile. North is approximately to the left and east down. The UV-luminous source discussed in Sect.~\ref{sec:uvbright} is visible just west of the centre in the F300X image.
}
\end{figure}

Inspection of the F300X image revealed a UV-bright source at a projected distance of about $0\farcs3$ or 1 pc west of the centre of EXT8. The source is quite distinct in the F300X image, but is much less conspicuous and suffers from blending with other nearby sources in F606W and F814W. To carry out photometry we first used the \texttt{ishape} task \citep{Larsen1999} in \texttt{BAOLAB} to subtract a PSF-convolved model of EXT8 from the images. Figure~\ref{fig:uvc} shows the F300X and F606W images of the central $2\farcs4\times2\farcs4$ of the cluster (left) and the residuals (right). We then obtained aperture photometry with the \texttt{phot} task in \texttt{IRAF}. 
In the F300X image we measured an apparent magnitude of $m_\mathrm{F300X}=21.87\pm0.04$ in an $r=5$~pixels ($0\farcs2$) aperture, or $M_\mathrm{F300X}=-3.19\pm0.04$ after applying an aperture correction of $-0.24$~mag. The source is thus more than 2 mag brighter in F300X than the other UV-bright stars discussed earlier (Sect.~\ref{sec:hbuv}).
In F606W it was difficult to obtain a meaningful measurement even on the model-subtracted image, owing to the very irregular background. For an $r=5$~pixels aperture we found $m_\mathrm{F606W}=24.48\pm1.36$, while a smaller ($r=3$~pixels) aperture yielded $m_\mathrm{F606W}=22.17\pm0.10$ and $m_\mathrm{F300X}=21.91\pm0.02$. Using the measurement in the smaller apertures we then find $M_\mathrm{F300X}-M_\mathrm{F606W}\simeq-0.45\pm0.10$ as a rough estimate of the colour. However, the $m_\mathrm{F606W}$ magnitude may be considered an upper limit, so the colour could be considerably bluer. 

Comparing with stars on the HB, which have $M_\mathrm{F300X}\approx+0.8$ and $L \approx 40 \, L_\odot$, the UV-bright source near the centre of EXT8 would have $L \approx 1600 \, L_\odot$. Again, the estimate of the bolometric luminosity $L$ is uncertain because of the large uncertainty on the colour. However, luminosities around $2000 \, L_\odot$ are typical for post-AGB stars \citep{deBoer1985,Bond2021} such as Barnard~29 in M13 \citep{Dixon2019}. Post-AGB stars evolve rapidly across the H--R diagram on time scales of $\sim10^5$~yr and are therefore relatively rare, with Galactic GCs typically hosting 0--2 such stars \citep{Moehler2019}. Finding a single post-AGB star candidate in EXT8 is therefore not entirely unexpected. It may be noted that the detailed evolution of post-AGB stars remains uncertain, in part because of their relative rarity. It is unlikely that any large number of such stars remain undiscovered in the Milky Way GC system, but the observations presented here demonstrate that a search for post-AGB stars in M31 GCs and other Local Group GCs should be quite feasible with space-based UV imaging. 

\subsection{Integrated properties and radial structure}
\label{sec:iprop}

To measure the integrated properties of EXT8 we used the \texttt{photutils} package \citep{Bradley2020} in \texttt{Python} to carry out aperture photometry on the drizzle-combined images (the \texttt{\_drc} files) in concentric apertures out to a radius of 12\arcsec, measuring the background between 16\arcsec\ and 20\arcsec\ from the cluster centre. These apertures were chosen to avoid the two bright stars at 13\arcsec\ and 22\arcsec\ from EXT8 (Fig.~\ref{fig:img}).
Within the 12\arcsec\ aperture, the total STMAG magnitudes are 
$m_\mathrm{F300X} = 16.48$, 
$m_\mathrm{F606W} = 15.58$ and $m_\mathrm{F814W} = 15.89$, with formal uncertainties $<0.01$~mag. 
This gives
$M_\mathrm{F300X} = -8.34$,
$M_\mathrm{F606W} = -9.06$ and $M_\mathrm{F814W} = -8.68$ for the assumed distance and extinction.
The UV-bright source discussed in Sect.~\ref{sec:uvbright} accounts for about 0.9\% of the cluster luminosity in F300X and an even lower fraction at optical wavelengths, and is therefore unlikely to have significantly affected the spectroscopic measurements of the integrated light.

We converted the WFC3 integrated photometry to standard Johnson--Cousins $V$ and $I$ magnitudes using the transformations in \citet{Harris2018}. This gives $V_0 = 15.28$, $M_V=-9.19$, and $(V-I)_0 = 0.75$, confirming that EXT8 is very blue for a GC as expected for its low spectroscopic metallicity.
For comparison, \citet{Mackey2019} found $(V-I)_0 = 0.79$ and $M_V = -9.28$, which makes EXT8 the bluest of the clusters included in their study.
The difference between the $(V-I)_0$ colour reported by \citet{Mackey2019} and that found here may be partly caused by uncertainties in the transformation from the WFC3 system to the Johnson--Cousins system, which is regarded  as particularly difficult for the F606W filter and is likely uncertain by $\simeq0.1$~mag near $(V-I)_0\simeq0.8$ \citep{Harris2018}.

Half of the counts in the F606W image are contained within a half-light radius of $R_h = 0\farcs92$ or about 3.5 pc, which is somewhat larger than the $R_h = 0\farcs73\pm0\farcs05$ ($2.8\pm0.2$~pc) measured on a ground-based image by \citet{Larsen2020}.
There is a slight trend with wavelength, with $R_h=0\farcs90$ in the F300X image and $R_h=0\farcs95$ in F814W. The formal errors on the $R_h$ measurements due to counting statistics are small, $<0\farcs01$. Systematic uncertainties are likely larger, but harder to estimate, and may be caused by uncertainties in the background subtraction, the centring of the apertures, and the behaviour of the profile beyond 12\arcsec, as discussed further below.
The relatively compact size of EXT8 follows the trend seen for other luminous GCs in the halo of M31 \citep{Huxor2014}.

We next used the \texttt{kfit2d} task in  \texttt{BAOLAB} to fit a 2-dimensional \citet{King1966} profile to the drizzled F606W image of EXT8. The background was measured within the same 16\arcsec--20\arcsec\ annulus used for the aperture photometry, and the fit was carried out within a radius of 12\arcsec.
The best fit was obtained for a model profile with a concentration parameter of $W_0=7.8$ and from this fit we found an ellipticity of $\epsilon=(1-b/a)=0.20\pm0.01$ (for minor/major axis ratio $b/a=0.80\pm0.01$), a core radius measured along the major axis of $r_c = 0\farcs238\pm0\farcs001$, and a projected half-light radius of $R_h=0\farcs97\pm0\farcs02$. The ellipticity is very similar to that measured from the ground-based image \citep[$\epsilon=0.19$;][]{Larsen2020}, while the half-light radius is again larger but agrees fairly well with the aperture photometry above. 

\begin{figure}
\includegraphics[width=89mm]{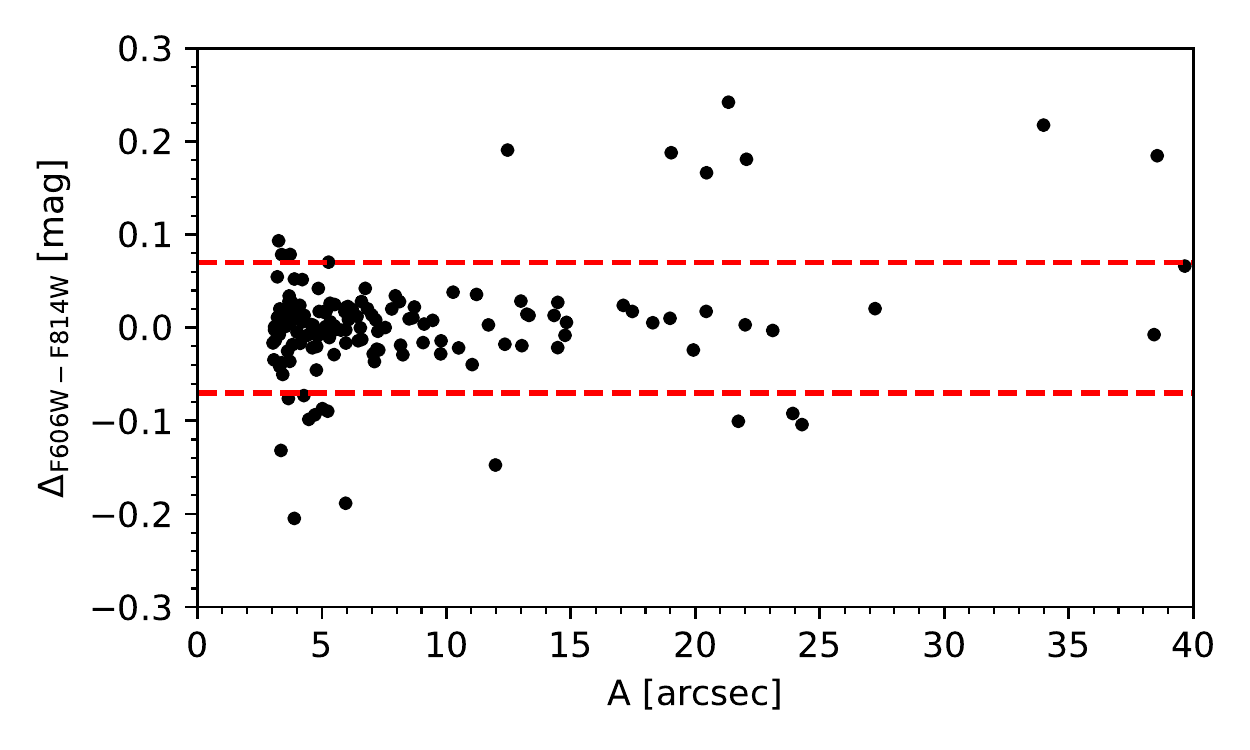}
\caption{\label{fig:dvi_r}Radial distribution of $\Delta_\mathrm{F606W-F814W}$ offsets for stars brighter than $m_\mathrm{F606W}=25$. The horizontal dashed lines demarcate the $\Delta_\mathrm{F606W-F814W}$ range used for selection of RGB stars in EXT8 (cf. Fig.~\ref{fig:cmds}). 
}
\end{figure}

\begin{figure}
\includegraphics[width=89mm]{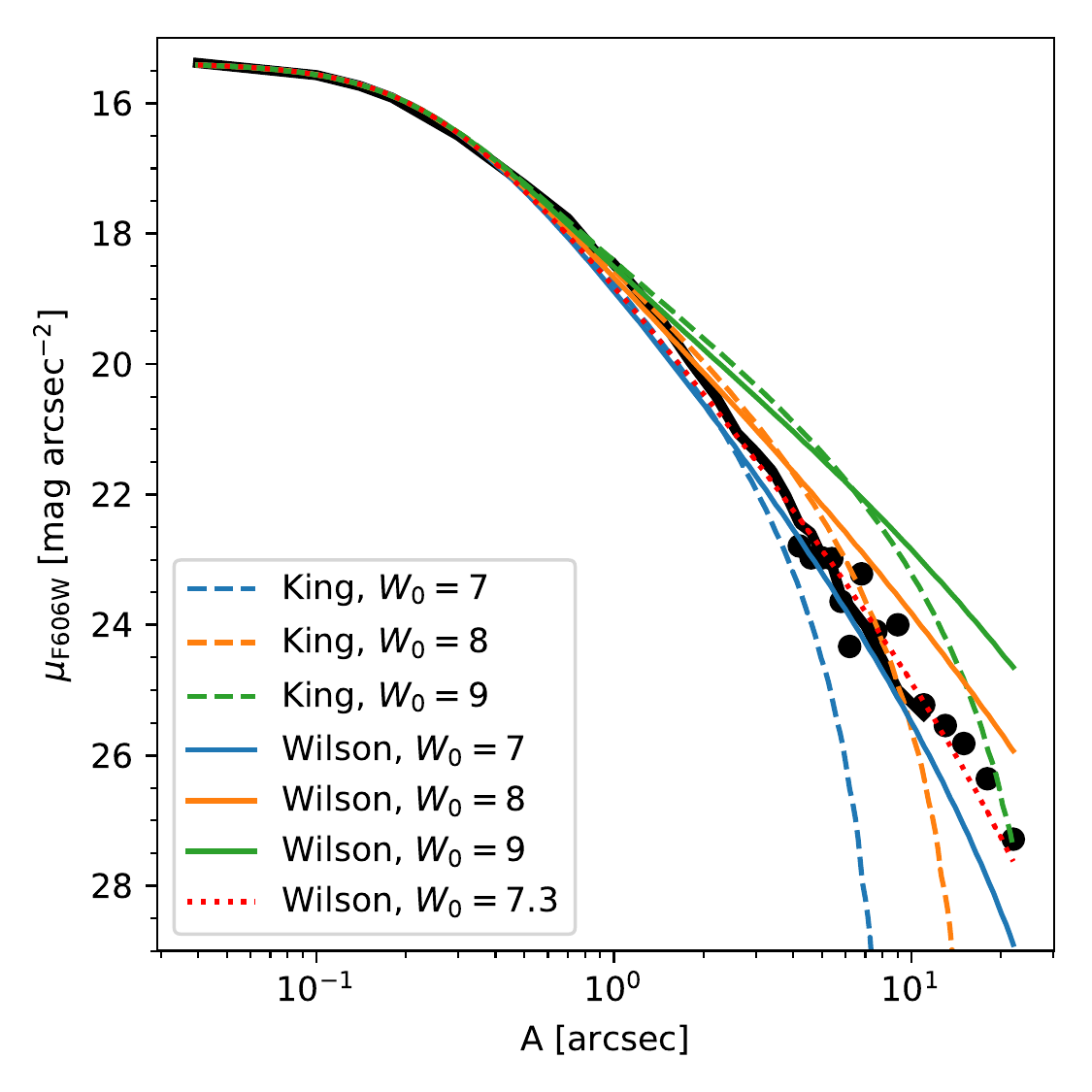}
\caption{\label{fig:kfit}The surface brightness profile of EXT8 and various model profiles, both shown versus semi-major axis $A$. Filled circles are based on star counts, the solid black curve on surface photometry. \citet{King1966} and \citet{Wilson1975} models are drawn with dashed and solid curves. 
The best-fitting model, a Wilson model with $W_0=7.3$, is plotted as a dotted curve.
}
\end{figure}

Beyond a radius of about 12\arcsec, surface photometry becomes difficult due to the low surface brightness of the cluster and the two bright foreground stars. However, we  used star counts to extend the profile out to about 25\arcsec. 
Figure~\ref{fig:dvi_r} shows the $\Delta_\mathrm{F606W-F814W}$ offsets as a function of the semi-major axis ($A$) of an ellipse drawn through each star. To calculate the semi-major axes
we adopted the ellipticity and orientation of EXT8 from the 2-D King profile fits. To improve the statistics we included stars to a slightly fainter magnitude limit than in Sect.~\ref{sec:zspread}, $m_\mathrm{F606W}=25.0$, which is still expected to give a clean and complete sample of EXT8 RGB stars (Fig.~\ref{fig:cmds}). The sequence of RGB stars, centred on $\Delta_\mathrm{F606W-F814W}=0$, can be traced out to a semi-major axis of $A=25\arcsec$, beyond which point very few stars with colours matching the EXT8 RGB stars are found.

In Fig.~\ref{fig:kfit} we combine the data from Fig.~\ref{fig:dvi_r} with surface photometry on the F606W image to show the surface brightness profile from the centre out to a semi-major axis of 25\arcsec . For the surface photometry we used \texttt{photutils} to measure the flux in elliptical annuli.
For the outermost regions we selected RGB stars with $\left|\Delta_\mathrm{F606W-F814W}\right| < 0.07$~mag (red dashed lines in Fig.~\ref{fig:dvi_r}) and scaled the density profile obtained from the number counts such that it matched the surface photometry in the region of overlap ($4\arcsec < A < 12\arcsec$).

Also shown in Fig.~\ref{fig:kfit} are single-mass \citet{King1966} and \citet{Wilson1975} model profiles with concentration parameters of $W_0 = 7$, 8, and 9. All model profiles have the same central surface brightness as the observed profile, $\mu_{F606W}(A=0) = 15.39$~mag~arcsec$^{-2}$ with a formal uncertainty less than 0.01~mag~arcsec$^{-2}$. The profiles were calculated with the \texttt{LIMEPY} package \citep{Gieles2015} and have been scaled such that all model profiles have the same projected half-width at half maximum as the observed profile, $\mathrm{HWHM} = 0\farcs23$ measured along the semi-major axis. The HWHM is very similar to the core radius parameter $r_c$ (also sometimes called the King radius) and has a similarly small formal uncertainty ($<0\farcs001$), but the conversion depends weakly on the concentration parameter. The core radii are between 0\farcs24 and 0\farcs25 for the profiles shown, similar to the value found from the 2-D fit. Because the radial profile analysis does not account for the PSF, the central surface brightness may be underestimated by 0.1-0.2 mag and the core radius may be overestimated by $\sim0\farcs02$ \citep{Chandar2001,Barmby2002,Larsen2002b}.

The various model profiles differ little in the central regions where they all fit the observed profile quite well. In the outer parts of the cluster, the \citet{King1966}-type profiles do not provide good fits, while better fits are obtained for the more extended \citet{Wilson1975} profiles. The best fit to the outermost parts is obtained for a $W_0\approx7.3$ Wilson profile, although such a profile has a slight deficit of light at intermediate radii (around 1\arcsec) compared to the observations. 
There is also a hint that the observed profile drops somewhat more steeply beyond $A=25\arcsec$, where extrapolation of the Wilson profile predicts an additional five stars. There is no evidence of an extended extra-tidal power-law envelope as observed around some Galactic GCs \citep{Kuzma2018}.

\section{Discussion}

For the most part, EXT8 resembles a normal globular cluster, the main exceptions being its extremely low metallicity and the unusually low magnesium abundance derived from the spectroscopic analysis \citep{Larsen2020}.
The CMD is fully consistent with a metal-poor, mono-metallic, old stellar population with a blue HB and a slightly bluer and steeper RGB than in the most metal-poor Galactic GCs. Nevertheless, the small intrinsic colour spread that is allowed on the RGB within the observational uncertainties could still translate to a substantial relative metallicity spread. While this is a natural consequence of the fact that RGB colours become less sensitive to metallicity at low metallicities, the exact dependence of RGB colour on metallicity is very model dependent. 
A relative metallicity dispersion of $\sim0.3$ dex, as obtained by scaling the empirical difference between the RGB colours of EXT8 and M92/M30 to the maximum allowed colour spread (Sect.~\ref{sec:zspread}), 
would still be similar to the largest spreads observed in Galactic GCs such as $\omega$~Cen and M54 \citep{Willman2012}, although much larger metallicity dispersions have been claimed in massive M31 GCs (\citealt{Meylan2001,Fuentes-Carrera2008}, but see also \citealt{Stephens2001}). Of course, the absolute metallicity spreads are in all these cases much larger than the allowable spread in EXT8, due to their much higher mean metallicities, and the limit on the metallicity spread in EXT8 is smaller than the spreads typically observed in metal-poor dwarf galaxies \citep{Willman2012}.

The exact age of EXT8 is not well constrained by the CMD, since it does not reach the MSTO and models predict a blue HB morphology even for GCs that are up to $\sim2$~Gyr younger than a canonical ``old'' GC at  $\mathrm{[Fe/H]}\simeq-2.5$ \citep{Catelan1993,Lee1994,Rey2001}. We are not aware of specific predictions at the even lower metallicity of EXT8, but it appears likely that its HB morphology would allow an even younger age. 
For GCs in the Milky Way, a considerable age spread has been inferred from variations in HB morphology as well as more direct age indicators such as the MSTO, with a tendency for younger clusters to be preferentially located at relatively large distances from the Galactic centre \citep{Searle1978,MarinFranch2009,Forbes2010}. Similar results have been found for GCs in M31 \citep{Perina2012,Mackey2013}. These age differences are most pronounced at $\mathrm{[Fe/H]}\ga-1.5$, where the ages of the younger GCs appear to fit chemical enrichment models for potential progenitor dwarf galaxies in which the GCs may have resided before being accreted onto the halos of the larger galaxies \citep{Kruijssen2019a,Forbes2020}. At metallicities $\mathrm{[Fe/H]}<-2$ there is less evidence for any significant age spread among GCs \citep{Dotter2011}, a result which appears consistent with more general constraints on age-metallicity relations \citep{Muratov2010,Bellstedt2020,Horta2021}. 
In all likelihood, EXT8 is at least as old as the oldest GCs observed in the Milky Way and elsewhere, although confirmation of this hypothesis must await deep photometry (for example with the \emph{James Webb Space Telescope}) that can reach below the MSTO. 

It might be argued that the HB morphology of EXT8 is \textit{unexpectedly} blue, given that the bluest HBs tend to be found in clusters like M10 and M13 with $\mathrm{[Fe/H]}\simeq-1.5$ and not generally in the most metal-poor GCs \citep{Renzini1983,Sandage1990a,Dotter2010}.
Furthermore, the general tendency is for GCs at larger galactocentric distances to exhibit redder HB morphologies than those in the inner halos both in the Milky Way and M31 \citep{Lee1994,Perina2012}. The very blue HB of EXT8 and its relatively large galactocentric distance may be contrary to this trend, although this must remain a tentative suggestion due to the uniquely low metallicity and corresponding lack of a suitable comparison sample.

Structurally, EXT8 resembles normal compact, luminous Galactic GCs, and it lacks the extended extra-tidal halos seen in some GCs that are suspected to be stripped nuclei \citep{DaCosta2015,Kuzma2018}. It is not unusual for the more extended \citet{Wilson1975}-type profiles to fit the outer regions of GCs better than the classical \citet{King1966} profiles  \citep{McLaughlin2005}, as we find for EXT8 -- both are relatively simple dynamical models which differ in their assumptions about the effect of a finite escape velocity on the distribution function \citep{Gieles2015}. While the ellipticity of $\epsilon=0.20$ indicates a noticeable degree of flattening, ellipticities up to 0.3 or more are observed among GCs in the Milky Way as well as other galaxies, with a tendency for relatively massive GCs to be more flattened
\citep{vandenBergh1984,Harris2002a,Gomez2006}. Hence, EXT8 is not particularly exceptional in this respect either. 
We note that although rotation in a GC could in principle provide clues about its origins, there has not been a clear connection established in Galactic GCs between ellipticity and rotation
\citep{Kamann2018,Sollima2019}.

It appears likely that the formation of EXT8 must be understood within the general context of GC formation scenarios. From the mass--metallicity relation in \citet{Choksi2018}, a metallicity of $\mathrm{[Fe/H]}=-2.9$ corresponds to a proto-galactic halo with a stellar mass of about $2\times10^4 \, M_\odot$ at a redshift of $z=5$ and about $10^5 \, M_\odot$ at $z=10$, which illustrates the fundamental difficulty of forming a GC this metal-poor and massive in a low-mass proto-galaxy, as envisioned in hierarchical galaxy formation scenarios \citep{Choksi2018,Kruijssen2019}. The difficulty is compounded if we consider that EXT8 probably did not form in isolation, although the inverse correlation between metallicity and GC specific frequency suggests that GCs might have formed more efficiently at lower metallicities \citep{Harris2002,Beasley2008,Lamers2017}.

Of course, the actual mass--metallicity relation in the low mass, high-$z$ regime is poorly constrained from observations, and it remains unclear how much of a challenge a single object like EXT8 poses. It would clearly be of substantial interest to better characterise the low-metallicity tail of the GC metallicity distribution and compare with simulations tailored to explore this part of parameter space. 
The existence of objects like EXT8 could inform models of galaxy enrichment in the extremely low metallicity regime \citep{Wheeler2019,Agertz2020}.

The fact that EXT8 stands out by its blue integrated colours suggests a route to identifying additional candidates, although colours are also sensitive to other parameters besides metallicity such as HB morphology, age, and variations in the MF \citep{Vanderbeke2013a}. According to the \citet{Harris1996} catalogue, the three metal-poor Galactic GCs discussed in Sect.~\ref{sec:cmds} have colours between $(V-I)_0 = 0.71$ (M15) and $(V-I)_0=0.85$ (M92) in spite of their very similar metallicities.
It may here be noted, in particular, that M15 is even bluer than EXT8, so a search for GC candidates with very blue colours is not guaranteed to turn up only extremely metal-poor objects. NGC~4147 has a similar integrated colour to EXT8, $(V-I)_0 = 0.76$, despite its significantly higher metallicity. Confirmation of candidate extremely metal-poor GCs will therefore require spectroscopic follow-up. The challenge here is that the metallic features become weak at such low metallicities, such that relatively high signal-to-noise spectra are needed for robust measurements. 

\section{Summary and conclusions}

We have presented an analysis of \textit{Hubble Space Telescope} observations of the extremely metal-poor globular cluster EXT8 in M31. The main conclusions are as follows:

\begin{itemize}
    \item The CMD of EXT8 looks as expected for a metal-poor, old stellar population, with a steep RGB and a blue HB, and unambiguously confirms that EXT8 is an old, metal-poor GC. 
    \item The colour of the upper EXT8 RGB is about 0.03--0.04 mag bluer in $M_\mathrm{F606W}-M_\mathrm{F814W}$ than the RGBs of the metal-poor Galactic GCs M92 and M30, but is similar in colour to the M15 RGB. The latter may however be more affected by uncertainties in the foreground reddening.
    \item The slope of the RGB is consistent with an extension of the relation between RGB slope and metallicity found by \citet{Sakari2015} to the lower metallicity of EXT8.
    \item The colour spread on the RGB is consistent with being caused entirely by observational uncertainties, although an intrinsic dispersion in $M_\mathrm{F606W}-M_\mathrm{F814W}$ of up to about 0.015~mag is allowed. 
    \item Translating the upper limit on the colour dispersion to a limit on the allowed metallicity spread is very model dependent. According to MIST isochrones, the maximum colour spread corresponds to about 0.2~dex in metallicity, but when using BaSTI isochrones a much larger spread of at least 0.7~dex is allowed. 
    \item The HB is located mainly on the blue side of the RR Lyrae gap and extends at least down to $M_\mathrm{F606W}=+3$. A further extension of the HB would be below our detection limit and thus cannot be excluded.  We quantify the HB morphology as $\mathrm{SMI} \ga 0.9$ and $\Delta(V-I) > 0.95$, which is comparable to the bluest HBs observed in Galactic GCs.
    \item Two stars located in or near the instability strip have a larger rms scatter in F606W and F814W than expected from their mean photometric errors and may be considered RR Lyrae candidates. 
    \item The CMD shows a few candidate UV-bright stars above the horizontal branch. These are the brightest stars in the $M_\mathrm{F300X}$ vs.\ $M_\mathrm{F300X}-M_\mathrm{F606W}$ CMD.
    An even brighter source is located near the centre and is a likely post-AGB star candidate. 
    \item The surface brightness profile is well fitted by a \citet{Wilson1975} model with a central surface brightness of $\mu_\mathrm{F606W,0}=15.2$~mag~arcsec$^{-2}$ (corrected for foreground extinction), an ellipticity $\epsilon=0.20$, and a core radius of $r_c=0\farcs25$ (about 0.93~pc) measured along the semi-major axis. There is no evidence of extra-tidal stars.
    \item From the RGB tip and the HB, we find that the distance modulus of EXT8 likely lies between $(m-M)_0 = 24.43\pm0.03$ and $24.57\pm0.05$. Depending on the distance of M31 itself, the
    true 3-D distance between EXT8 and the centre of M31 could be between the projected value (27~kpc) and up to about 70~kpc.
\end{itemize}

We conclude that EXT8 has properties consistent with it being a ``normal'', albeit very metal-poor GC. While its combination of relatively high mass and very low metallicity are challenging to explain in the context of GC formation theories operating within the hierarchical galaxy assembly paradigm, a more systematic search for similar objects would be highly desirable. 

The results of this work reaffirm the dual roles that GCs can play in bridging stellar and (extra-)galactic astrophysics. Finding more objects like EXT8 would not only help constrain galaxy formation theories, but would also provide valuable input for the calibration of stellar model properties such as the RGB colour, HB morphology, and post-AGB evolution at low metallicities.

\begin{acknowledgements}

AJR was supported by National Science Foundation grant AST-1616710 and as a Research Corporation for Science Advancement Cottrell Scholar.
This research is based on observations with the NASA/ESA {\it Hubble Space Telescope}
obtained at the Space Telescope Science
Institute, which is operated by the Association of Universities for
Research in Astronomy, Incorporated, under NASA contract NAS5-26555. Support for Program number HST-GO-16459 was
provided through a grant from the STScI under NASA contract NAS5-26555.
This research has made use of the NASA/IPAC Extragalactic Database (NED),
which is operated by the Jet Propulsion Laboratory, California Institute of Technology, under contract with the National Aeronautics and Space Administration. This research has made use of NASA’s Astrophysics Data System Bibliographic Services.
We thank the anonymous for very quickly providing a report that nevertheless contained several useful suggestions that helped us improve the manuscript. We also thank Annette Ferguson, Silvia Martocchia, and Anastasia Gvozdenko for useful comments and discussion. 

\end{acknowledgements}

\bibliographystyle{aa}
\bibliography{refs.bib}

\appendix

\section{Photometry}

\begin{table*}
\caption{F606W/F814W photometry. Only the first few rows are reproduced here; the full table is available at the CDS.}
\label{tab:photometry}
\centering
\begin{tabular}{rcccccccccc}
\hline\hline
ID & $X$ & $Y$ & \multicolumn{3}{c}{F606W} & \multicolumn{3}{c}{F814W} \\
& & & mag & error & rms & mag & error & rms \\ \hline
     1 &   52.959 &  52.579 & 26.056 & 0.120 & 0.152 & 26.570 & 0.117 & 0.126 \\
     5 & 4047.204 &  52.729 & 26.053 & 0.121 & 0.125 & 26.559 & 0.114 & 0.202 \\
    24 & 3961.471 &  52.602 & 27.827 & 0.138 & 0.095 & 28.231 & 0.165 & 0.233 \\
    36 & 3945.614 &  53.102 & 27.980 & 0.099 & 0.272 & 27.952 & 0.124 & 0.259 \\
    40 &  623.995 &  53.526 & 27.561 & 0.079 & 0.182 & 27.610 & 0.079 & 0.229 \\
\hline
\end{tabular}
\tablefoot{$X$ and $Y$ are the coordinates on the WFC3 UVIS2 F606W image.
}
\end{table*}

\begin{table*}
\caption{F300X/F606W/F814W photometry. Only the first few rows are reproduced here; the full table is available at the CDS.}
\label{tab:uviphotometry}
\centering
\begin{tabular}{rccccccccccc}
\hline\hline
ID & $X$ & $Y$ & \multicolumn{3}{c}{F300X} & \multicolumn{3}{c}{F606W} & \multicolumn{3}{c}{F814W} \\
& & & mag & error & chi & mag & error & chi & mag & error & chi \\ \hline
     4  & 23.631 & 3.019 & 26.771 & 0.629 & 3.53 & 29.283 & 0.702 & 2.01 & 28.566 &   0.310 & 1.98 \\
     5  & 30.314 & 3.110 & 26.903 & 0.340 & 1.71 & 26.444 & 0.331 & 6.21 & 29.535 &  0.640 & 1.76 \\
     7  & 38.383 & 2.947 & 27.393 & 0.675 & 2.31 & 28.756 & 0.361 & 1.71 & 29.325 &  0.469 & 1.56 \\
    15  &  2.689 & 8.221 & 22.539 & 0.249 & 16.75 & 30.129 & 1.386 & 1.74 & 28.854 &   0.569 & 2.74 \\
    17  & 17.889   & 10.396 & 27.261 & 0.514 & 1.91 & 28.322 & 0.217 & 1.36 & 30.318 &  0.954 & 1.30 \\
\hline
\end{tabular}
\tablefoot{$X$ and $Y$ are the coordinates on the drizzled WFC3 F300X image.
}
\end{table*}

\end{document}